\documentstyle[seceq,epsf,mbf,wrapft]{ptptex}
\def\ba{{\mbf a}}
\def\bb{{\mbf b}}
\def\bp{{\mbf p}}
\def\bq{{\mbf q}}
\def\bk{{\mbf k}}
\def\bn{{\mbf n}}
\def\br{{\mbf r}}

\def\bx{{\mbf x}}
\def\bz{{\mbf z}}

\def\bfsigma{{\mbf \sigma}}

\def\bftau{{\mbf \tau}}

\def\bA{{\mbf A}} 
\def\bS{{\mbf S}}
\def\bK{{\mbf K}}

\def\CM{{\cal M}}
\def\bP{{\mbf P}}
\def\bR{{\mbf R}}

\def\bV{{\mbf V}}
\def\bX{{\mbf X}}
\def\bW{{\mbf W}}
\def\CP{{\cal P}}
\def\CT{{\cal T}}
\def\CO{{\cal O}}

\def\CA{{\cal A}}

\def\eq#1{Eq.\,(\ref{#1})}

\def\lam#1{{#1 \over \lambda}}
\def\lab#1{{#1 \over \lambda^2}}
\def\ttau{\bftau_1 \cdot \bftau_2}
\def\ttaup{(1+\ttau) P_F}
\def\SM3{\Sigma N (3/2)}
\def\SN1{\Sigma N (1/2)}

\def\TS1{\hbox{}^3S_1}
\def\TD1{\hbox{}^3D_1}
\def\CA{{\cal A}}
\def\CM{{\cal M}}
\def\CP{{\cal P}}
\def\CT{{\cal T}}
\def\CO{{\cal O}}

\def\CA{{\cal A}}

\def\CY{{\cal Y}}
\def\CZ{{\cal Z}}

\def\aagp{{\alpha \alpha^\prime}}

\def\lesssim{\begin{array}{c} < \\[-3mm] \sim \\ \end{array}}

\def\bfsigma{{\mbf \sigma}}
\def\bftau{{\mbf \tau}}

\def\li{{1 \over \lambda}}

\def\eq#1{Eq.\,(\ref{#1})}

\def\lam#1{{#1 \over \lambda}}
\def\lab#1{{#1 \over \lambda^2}}
\def\ttau{\bftau_1 \cdot \bftau_2}
\def\ttaup{(1+\ttau) P_F}
\def\ssigma{\bfsigma_1 \cdot \bfsigma_2}
\notypesetlogo  %comment in if to eliminate PTPTeX logo
\preprintnumber{%To give preprint# at top right corner when [preprint]
KUNS-14?? \\ June 1998}
\title{Lippmann-Schwinger Resonating-Group Formalism
for $\mbf{NN}$ and $\mbf{YN}$ Interactions 
in an $\mbf{SU_6}$ Quark Model}

\author{
Yoshikazu {\sc Fujiwara}, Michio {\sc Kohno}$^{*}$,
Tadashi {\sc Fujita},
Choki {\sc Nakamoto}$^{**}$
and Yasuyuki {\sc Suzuki}$^{***}$
}

\inst{
Department of Physics, Kyoto University,
Kyoto 606-8502 \\
$\hbox{}^{*}$Physics Division, Kyushu Dental College,
Kitakyushu 803-8580, Japan \\
$\hbox{}^{**}$Suzuka National College of Technology,
Suzuka 510-0294, Japan \\
$\hbox{}^{***}$Department of Physics, Niigata University,
Niigata 950-2181, Japan}
\recdate{%      %Editorial Office will fill in this.
\today
}
\abst{%         %this abstract is neglected when [addenda] or [errata]
We formulate a Lippmann-Schwinger-type resonating-group equation
to calculate invariant amplitudes of the quark-model
baryon-baryon interaction.
When applied to our recent $SU_6$ quark model for the nucleon-nucleon
and hyperon-nucleon interactions, this technique
yields very accurate phase-shift parameters for all partial waves
up to the energies of several GeV. The technique also has a merit
of a straightforward extension to the $G$-matrix equation.
A new analytic method is proposed to calculate the quark-exchange
Born kernel for the momentum-dependent two-body interaction.
The partial-wave decomposition in the momentum representation
is carried out numerically.
The invariant amplitudes are then used to calculate
single-nucleon potentials in normal nuclear matter
for high incident momenta $q_1 \geq 3~\hbox{fm}^{-1}$,
in which the so-called $t^{eff}\rho$ prescription
is found to be a good approximation
to the single-particle potentials directly calculated
in the lowest-order Brueckner theory.
}

\begin{document}

\maketitle

\section{Introduction}

Though the quantum chromodynamics (QCD) is believed
to be the fundamental theory of the strong interaction,
it is still too difficult to apply it directly to two-baryon systems. 
At this stage a number of effective models have been
proposed to understand the nucleon-nucleon ($NN$) and
hyperon-nucleon ($YN$) interactions from basic
elements of quarks and gluons. \cite{CL79}
Among them the non-relativistic quark model
has a unique feature that it enables us to take
full account of a dynamical motion of the two composite baryons
within a framework of the resonating-group method (RGM). \cite{WS84}
The model describes confinement with a phenomenological
potential and uses quark-quark ($qq$) residual interactions
consisting of a color analogue of the Fermi-Breit (FB) interaction.
In the last several years, it was found that a proper incorporation
of the meson-exchange effect is essential
to make such a model realistic for the description
of the $NN$ and $YN$ interactions.
\cite{NA95,FU95,FU96a,FU96b,FJ98,YU95,ZH97,YA98}

We have recently achieved a simultaneous description of the $NN$
and $YN$ interactions in the RGM formulation
of the spin-flavor $SU_6$ quark
model. \cite{NA95,FU95,FU96a,FU96b,FJ98}
In this model the meson-exchange effect
of scalar (S) and pseudo-scalar (PS) meson nonets
is incorporated in the quark Hamiltonian
in the form of effective meson-exchange potentials (EMEP) acting
between quarks.
The flavor symmetry breaking for the $YN$ system is explicitly
introduced through the quark-mass dependence of the Hamiltonian,
as well as the flavor dependence of the exchanged meson masses.
An advantage of introducing the EMEP at the quark level lies
in the stringent relationship of the flavor dependence,
which is revealed in the various pieces
of the $NN$ and $YN$ interactions.
In this way we can utilize our rich knowledge of the $NN$
interaction to minimize the ambiguity of model parameters, 
which originates from the scarcity of the present experimental data
for the $YN$ interaction.

We already have three different versions called
RGM-F \cite{NA95,FU95}, FSS \cite{FU96a,FU96b,FJ98} and
RGM-H \cite{FU96b,FJ98}, differing in the treatment of the EMEP.
The model called RGM-F introduces,
besides the central force of the S-meson nonet,
only the tensor component generated
from the $\pi$- and $K$-meson exchanges,
and uses some approximations in evaluating the spin-flavor
factors of the quark-exchange RGM kernel.
On the other hand, FSS and RGM-H calculate the spin-flavor factors
explicitly at the quark level, and include
the spin-spin terms originating
from all members of the PS-meson nonet.
The $SU_3$ relation of the coupling constants emerges as a natural
consequence of the $SU_6$ quark model. For S-mesons,
the $F/(F+D)$ ratio turns out
to take the $SU_6$ value of purely electric type.
This is too restrictive to reproduce existing experimental data
for the low-energy $YN$ cross sections.
We relax this restriction in two ways; one is to change
the mixing angle of the flavor-singlet and octet scalar mesons
only for the $\Sigma N (I=3/2)$ channel,
and the other is to employ the same approximation as RGM-F solely
for the isoscalar S-mesons, $\epsilon$ and $S^*$.
We call these models FSS and RGM-H, respectively. Predictions
of these two models are not very different except for
the roles of the $LS^{(-)}$ force
in the $\Lambda N$ - $\Sigma N(I=1/2)$ coupled-channel system.
The $SU_3$ parameters of EMEP, S-meson masses,
and the quark-model parameters are determined to fit the $NN$ $S$-wave
and $P$-wave phase shifts under the constraint that
the deuteron binding energy and the $\hbox{}^1S_0$ scattering
length are properly reproduced. The low-energy cross section data
for $YN$ scattering are also employed to fix the parameters in the
strangeness sector, especially the up-down to strange quark-mass
ratio $\lambda=m_s/m_{ud}$. The reader is referred to
the original papers \cite{FU96b,FJ98} for a full account
of the models and the model parameters.

So far we have solved the coupled-channel (CC) RGM equation
in the improved variational method developed by Kamimura \cite{KA77}.
In this method each Gaussian basis function is smoothly
connected to the positive- or negative-energy
asymptotic waves, which are obtained by numerically solving
a ``local'' CC Schr{\"o}dinger equation consisting of the long-range
one-pion tensor force and the other EMEP.
Although this technique gives accurate results
at laboratory energies up to about 300 MeV,
it seems almost inaccessible to higher energies
due to the rapid oscillation of the relative wave functions.
In this paper we formulate an alternative method
to solve the CC RGM equation in the momentum representation;
namely, we derive a Lippmann-Schwinger-type RGM equation
which we call an LS-RGM equation.\footnote{The idea to solve
the RGM equation in the momentum representation, in order to avoid
the rapid oscillation of relative wave functions at higher energies,
is not new. See, for example, Ref.\,\citen{KU93}.}
In this method all the necessary Born
amplitudes (or the Born kernel) for the
quark-exchange kernel are analytically derived
by using a new transformation formula,
which is specifically developed for momentum-dependent
two-body interactions acting between quarks.
The partial wave decomposition of the Born kernel is
carried out numerically in the Gauss-Legendre integral
quadrature. The LS-RGM equation is then solved
by using the techniques developed by Noyes \cite{NO65} and
Kowalski \cite{KO65}.
Although this method requires more CPU time than the variational
method, it gives very stable and accurate results
over a wide energy range.
Since we first calculate the Born amplitudes of the RGM kernel,
it is almost straightforward to proceed
to the $G$-matrix calculation. \cite{KO99}

As an application of the present formalism, we discuss
single-particle (s.p.) potentials in normal nuclear matter.
This application is motivated by the $G$-matrix calculation 
of the $NN$ and $YN$ interactions in the
continuous prescription of intermediate spectra \cite{KO99}.
The s.p. potentials of the nucleon and hyperons predicted
by the model FSS have a flaw that they are too attractive
in the momentum region $q_1=5 \sim 20~\hbox{fm}^{-1}$. 
We will show that this particular feature of FSS
is related to the ill-behavior of the spin-independent central
invariant amplitude at the forward angles, which is traced
back to too simple S-meson exchange EMEP in this model.
We analyze this problem by using an approximation
of the s.p. potential in the asymptotic momentum region,
in terms of the $T$-matrix solution of the LS-RGM equation.
This technique is sometimes called the $t^{eff}\rho$ prescription.
We find that this procedure gives a fairly good approximation
of the s.p. potentials predicted by the $G$-matrix calculation,
even for such a small momentum as $q_1 \sim 3~\hbox{fm}^{-1}$  
($T_{lab} \sim 200~\hbox{MeV}$), as long as the real part of the
s.p. potential is concerned.

In the next section we start from the standard RGM equation
for the $(3q)$-$(3q)$ system and derive the LS-RGM equation
in the momentum space. Appendix A gives a convenient
formula to calculate the Born kernel of the quark-exchange
RGM kernel. The formula is especially useful
in the case of the momentum-dependent $qq$ force.
The Born kernel of FSS is explicitly
given in $\S$\,2.2 and in Appendix B.
Since the flavor dependence of the
spin-flavor factors plays an essential role in the application 
of the present formalism to the $NN$ and $YN$ interactions,
the structure of the Born kernel and the scattering
amplitudes is carefully described in $\S$\,2.3 in terms
of the Pauli-spinor invariants. 
Some useful expressions of the partial wave decomposition
of the Born kernel and the invariant amplitudes
are given in Appendices C and D, respectively.
The $t^{eff}\rho$ prescription is derived
in $\S$\,2.4 as a method to calculate the asymptotic behavior
of the s.p. potentials in the $G$-matrix calculation.
In $\S$\,3 we compare the phase-shift parameters
obtained by the LS-RGM formalism with those by the improved
variational method. The system we choose as an
example is the most complicated system
of the $\Lambda N$ - $\Sigma N (I=1/2)$ channel coupling.
The invariant amplitudes and the scattering observables
of the $NN$ system in the intermediate energies
between $T_{lab}=400 \sim 800~\hbox{MeV}$ are discussed in $\S$\,4.
The s.p. potential of the nucleon in normal nuclear matter
is calculated in $\S$\,5, by using the $t^{eff}\rho$ prescription.
It is shown that the behavior of the spin-independent central
invariant amplitude at the forward angles is related to
the attractive behavior of the s.p. potentials
around $q_1=5 \sim 20~\hbox{fm}^{-1}$ region.
Also shown is a preliminary result of s.p. potentials
in an improved model,
which incorporates momentum-dependent higher-order terms
of the S and vector (V) meson EMEP. 
Finally, $\S$\,6 is devoted to a brief summary.

\newpage

\section{Formulation}

\subsection{LS-RGM equation}

As mentioned in the Introduction,
FSS \cite{FU96a,FU96b,FJ98}, RGM-H \cite{FU96b,FJ98}
and their preceding version RGM-F \cite{NA95,FU95} are
formulated in the $(3q)$-$(3q)$ RGM
applied to the system of two $(0s)^3$ clusters. 
The $qq$ interaction is composed of the full FB interaction
with explicit quark-mass dependence, a simple confinement
potential of quadratic power law and EMEP acting between quarks. 
The RGM equation for the parity-projected relative
wave function $\chi^\pi_\alpha({\bf R})$ is derived
from the variational
principle $\langle\delta\Psi|E-H|\Psi\rangle=0$, and it reads as 
\cite{FU96b} 
\begin{equation}
\left[~\varepsilon_\alpha + {\hbar^2 \over 2\mu_\alpha}
\left({\partial \over \partial {\bR}} \right)^2~\right]
\chi^\pi_\alpha({\bR})=\sum_{\alpha^\prime} \int d {\bR}^\prime
~G_\aagp({\bR}, {\bR}^\prime; E)
~\chi^\pi_{\alpha^\prime}({\bR}^\prime)\ \ ,
\label{fm1}
\end{equation}
where $G_{\alpha \alpha^\prime}({\bR}, {\bR}^\prime; E)$ is
composed of various pieces of the interaction kernels
as well as the direct potentials of EMEP:
\begin{eqnarray}
& & G_\aagp({\bR}, {\bR}^\prime; E)
=\delta({\bR}-{\bR}^\prime) \left[
\sum_\beta V^{(CN)\beta}_{\alpha \alpha^\prime \,D}({\bR})
+\sum_\beta V^{(SN)\beta}_{\alpha \alpha^\prime \,D}({\bR})
\right. \nonumber \\
& & \left. +\sum_\beta V^{(TN)\beta}_{\alpha \alpha^\prime \,D}({\bR})
~(S_{12})_{\alpha \alpha^\prime}~\right]
+ \sum_\Omega \CM^{(\Omega)}_\aagp ({\bR}, {\bR}^\prime)
-\varepsilon_\alpha~\CM^N_\aagp({\bR}, {\bR}^\prime)\ \ .
\label{fm2}
\end{eqnarray}
The subscript $\alpha$ stands for a set of quantum
numbers of the channel wave function; $\alpha=
\left[1/2(11)\,a_1, 1/2(11)a_2 \right]$ $SS_zYII_z; \CP$,
where $1/2(11) a$ is the spin and $SU_3$ quantum number
in the Elliott notation $(\lambda \mu)$, $a(=YI)$ is the flavor label
of the octet baryons ($N=1(1/2),~\Lambda=00,
~\Sigma=01$ and $\Xi=-1(1/2)$),
and $\CP$ is the flavor-exchange phase. \cite{NA95}
In the $NN$ system with $a_1a_2=NN$, $\CP$ becomes redundant
since it is uniquely determined by the isospin as $\CP=(-1)^{1-I}$.
The relative energy $\varepsilon_\alpha$ in the
channel $\alpha$ is related to the total energy $E$ of the system
through $\varepsilon_\alpha=E-E^{int}_\alpha$ with $E^{int}_\alpha
=E^{int}_{a_1}+E^{int}_{a_2}$.
In \eq{fm2} the summation over $\Omega$ for the exchange
kernel $\CM_{\alpha\alpha'}^{(\Omega)}$ involves
not only the exchange kinetic-energy ($K$) term, but also
various pieces of the FB interaction, as well as several
components of EMEP. The FB interaction involves
the color-Coulombic ($CC$) piece, the momentum-dependent Breit
retardation ($MC$) piece, the color-magnetic ($GC$) piece,
the symmetric $LS$ ($sLS$) piece, the antisymmetric $LS$ ($aLS$) piece,
and the tensor ($T$) piece, The EMEP contains 
the central ($CN$) component from the S-mesons,
and the spin-spin ($SN$) and tensor ($TN$) terms
originating from the PS mesons.
The contribution from a 
particular meson exchange is denoted by $\beta$ 
in \eq{fm2} only for the direct part of 
$G_{\alpha \alpha^\prime}({\bR}, {\bR}^\prime; E)$.
The explicit form of these $qq$ forces is given
in Refs.\,\citen{FU96b} and \citen{SU84},
and the corresponding basic Born kernel defined through
\begin{eqnarray}
M^B_\aagp (\bq_f, \bq_i; E)
& = & \langle\,e^{i \bq_f \cdot {\bR}}\,\vert
\,G_\aagp ({\bR}, {\bR}^\prime; E)
\,\vert\,e^{i \bq_{\,i}\cdot {{\bR}}^\prime} \rangle \nonumber \\
& = & \langle\,e^{i \bq_f \cdot {\bR}} \eta_\alpha^{SF}\,\vert
\,G({\bR}, {\bR}^\prime; E)\,\vert\,e^{i \bq_{\,i}\cdot
{{\bR}}^\prime} \eta_{\alpha^\prime}^{SF} \rangle
\label{fm3}
\end{eqnarray}
is given in the next subsection and Appendix B.
Here $\eta_\alpha^{SF}$ is the spin-flavor wave function
at the baryon level, which is defined
in Eq.\,(2.9) of Ref.\,\citen{FU97}.

To start with, we use the well-known Green function to
convert the RGM equation \eq{fm1} to an integral equation 
which has a parity-projected incident plane wave
in the $\alpha$ channel:
\begin{eqnarray}
& & \chi^\pi_{\gamma \alpha}({\bR},\bk_\alpha)
=\delta_{\gamma, \alpha}~\left[~e^{i\bk_\alpha \cdot {\bR}}
+(-1)^{S_\alpha}\CP_\alpha~e^{-i\bk_\alpha \cdot {\bR}} \right]
\nonumber \\
& & -{1 \over 4\pi}{2\mu_\gamma \over \hbar^2}
\int d {\bR}^\prime~{e^{ik_\gamma |{\bR}-{\bR}^\prime|} \over
|{\bR}-{\bR}^\prime|} \sum_\beta \int d {\bR}^{\prime \prime}
~G_{\gamma \beta}({\bR}^\prime, {\bR}^{\prime \prime}; E)
~\chi^\pi_{\beta \alpha}({\bR}^{\prime \prime},\bk_\alpha)\ \ . 
\label{fm4}
\end{eqnarray}
Here $\pi=(-1)^{S_\alpha}\CP_\alpha
=(-1)^{S_\beta}\CP_\beta=(-1)^{S_\gamma}\CP_\gamma$, because of 
the parity conservation.
The asymptotic wave of \eq{fm4} with $R \rightarrow \infty$ is
given by
\begin{eqnarray}
& & \chi^\pi_{\gamma \alpha}(\bR,\bk_\alpha)
\sim \delta_{\gamma, \alpha}~\left[~e^{i\bk_\alpha \cdot \bR}
+(-1)^{S_\alpha}\CP_\alpha~e^{-i\bk_\alpha \cdot \bR} \right]
\nonumber \\
& & +{1 \over k_\alpha}\sqrt{{v_\alpha \over v_\gamma}}
{e^{ik_\gamma R} \over R}
~\left[~M_{\gamma \alpha}(\bk_\gamma, \bk_\alpha; E)
+(-1)^{S_\alpha}\CP_\alpha
~M_{\gamma \alpha}(\bk_\gamma, -\bk_\alpha; E)~\right]\ ,
\label{fm5}
\end{eqnarray}
with $\bk_\gamma=k_\gamma \widehat{\bR}$ and
\begin{eqnarray}
& & M_{\gamma \alpha}(\bk_\gamma, \bk_\alpha; E)
=-{\sqrt{\mu_\gamma \mu_\alpha k_\gamma
k_\alpha} \over 4\pi \hbar^2} \nonumber \\
& & \times \sum_\beta \int d \bR \int d \bR^\prime
~e^{-i\bk_\gamma \cdot \bR}
~G_{\gamma \beta}(\bR, \bR^\prime; E)
~\chi^\pi_{\beta \alpha}(\bR^\prime,\bk_\alpha)\ . 
\label{fm6}
\end{eqnarray}
The Born amplitude is obtained by approximating
\begin{equation}
\chi^\pi_{\beta \alpha}(\bR,\bk_\alpha)
\sim \delta_{\beta, \alpha}~\left[~e^{i\bk_\alpha \cdot \bR}
+(-1)^{S_\alpha}\CP_\alpha
~e^{-i\bk_\alpha \cdot \bR} \right]
\label{fm7}
\end{equation}
in \eq{fm6}.
Though ${\bf k}_{\alpha}$ and ${\bf k}_{\gamma}$ are
related to the total energy $E$ by the on-shell condition
\begin{equation}
%\bk^2=\bk_\gamma^2\ \ , \qquad \bk^{\prime 2}=\bk_\alpha^2\ \ ,
E=E_\gamma^{int}+{\hbar^2 \over 2 \mu_\gamma}
\bk_\gamma^2=E_\alpha^{int}+{\hbar^2 \over 2 \mu_\alpha}
\bk_\alpha^2\ \ ,
\label{fm8}
\end{equation}
it is convenient to relax this condition in order to define a more 
general Born amplitude. 
Namely, $\bk_\gamma$ and $\bk_\alpha$ are denoted 
by $\bk$ and $\bk^{\prime}$, respectively, in what follows.
Then the Born amplitude reads
as\footnote{In Eqs.\,(\ref{fm9}), (\ref{fm20}) and (\ref{fm22}),
the relative energy $\varepsilon_\gamma$ in
the basic Born kernel (in the prior form)
should be calculated from the total
energy $E$ through $\varepsilon_\gamma=E-E_\gamma^{int}$.}
\begin{equation}
M^{\hbox{Born}}_{\gamma \alpha}(\bk, \bk^\prime; E)
=-{\sqrt{\mu_\gamma \mu_\alpha k
k^\prime} \over 4\pi \hbar^2} \left[
~M^B_{\gamma \alpha}(\bk, \bk^\prime; E)
+(-1)^{S_\alpha}\CP_\alpha
~M^B_{\gamma \alpha}(\bk, -\bk^\prime; E)~\right]\ \ .
\label{fm9}
\end{equation}
The Born amplitude \eq{fm9} has a high degree of symmetries
which can be directly derived from those of \eq{fm3}.
First $M^{\hbox{Born}}_{\gamma \alpha}(\bk, \bk^\prime; E)$ is
a function of $\bk^2$, $\bk^{\prime 2}$ and $\cos \theta
=\widehat{\bk}\cdot \widehat{\bk}^\prime$ with real coefficients,
and satisfies the symmetry
\begin{equation}
M^{\hbox{Born}}_{\gamma \alpha}(\bk, \bk^\prime; E)
=M^{\hbox{Born}}_{\gamma \alpha}(\bk, \bk^\prime; E)^*
=M^{\hbox{Born}}_{\alpha \gamma}(\bk^\prime, \bk; E)\ \ .
\label{fm10}
\end{equation}
The parity conservation implies
\begin{eqnarray}
M^{\hbox{Born}}_{\gamma \alpha}(\bk, \bk^\prime; E)
& = & M^{\hbox{Born}}_{\gamma \alpha}(-\bk, -\bk^\prime; E)
\nonumber \\
& = & (-1)^{S_\gamma}\CP_\gamma
~M^{\hbox{Born}}_{\gamma \alpha}(-\bk, \bk^\prime; E) \nonumber \\
& = & (-1)^{S_\alpha}\CP_\alpha
~M^{\hbox{Born}}_{\gamma \alpha}(\bk, -\bk^\prime; E)\ \ ,
\label{fm11}
\end{eqnarray}
with $(-1)^{S_\gamma}\CP_\gamma=(-1)^{S_\alpha}\CP_\alpha=\pi$.

We now move to the momentum representation in \eq{fm4} through
\begin{equation}
\chi^\pi_{\gamma \alpha}(\bR,\bk_\alpha)
=\int d \bk~e^{i\bk \cdot \bR}~\widehat{\chi}^\pi_{\gamma \alpha}
(\bk,\bk_\alpha)\ \ .
\label{fm13}
\end{equation}
Then \eq{fm6} is expressed as
\begin{equation}
M_{\gamma \alpha}(\bk_\gamma, \bk_\alpha; E)
=-{\sqrt{\mu_\gamma \mu_\alpha k_\gamma
k_\alpha} \over 4\pi \hbar^2} \sum_\beta \int d \bk^\prime
~V_{\gamma \beta}(\bk_\gamma, \bk^\prime; E)
~\widehat{\chi}^\pi_{\beta \alpha}(\bk^\prime,\bk_\alpha)\ \ , 
\label{fm14}
\end{equation}
where the Born kernel $V_{\gamma \beta}(\bk, \bk^\prime; E)$ defined
by
\begin{equation}
V_{\gamma \beta}(\bk, \bk^\prime; E)
={1 \over 2} \left[~M^B_{\gamma \beta}(\bk, \bk^\prime; E)
+(-1)^{S_\beta}\CP_\beta
~M^B_{\gamma \beta}(\bk, -\bk^\prime; E)~\right]
\label{fm15}
\end{equation}
is related to the Born amplitude through
\begin{equation}
M^{\hbox{Born}}_{\gamma \alpha}(\bk, \bk^\prime; E)
=-{\sqrt{\mu_\gamma \mu_\alpha k_\gamma
k_\alpha} \over 2\pi \hbar^2}
~V_{\gamma \alpha}(\bk, \bk^\prime; E)
\label{fm16}
\end{equation}
We write \eq{fm14} as
\begin{equation}
M_{\gamma \alpha}(\bk_\gamma, \bk_\alpha; E)
=-{\sqrt{\mu_\gamma \mu_\alpha k_\gamma
k_\alpha} \over 2\pi \hbar^2}
T_{\gamma \alpha}(\bk_\gamma, \bk_\alpha; E)
\label{fm17}
\end{equation}
and define the on-shell $T$-matrix by
\begin{equation}
T_{\gamma \alpha}(\bk_\gamma, \bk_\alpha; E)
={1 \over 2} \sum_\beta \int d \bk^\prime
~V_{\gamma \beta}(\bk_\gamma, \bk^\prime; E)
~\widehat{\chi}^\pi_{\beta \alpha}(\bk^\prime,\bk_\alpha)\ \ .
\label{fm18}
\end{equation}
Then the momentum representation of \eq{fm4} reads as
\begin{eqnarray}
& & \widehat{\chi}^\pi_{\gamma \alpha}(\bk,\bk_\alpha)
=\delta_{\gamma, \alpha}~\left[~\delta(\bk-\bk_\alpha)
+(-1)^{S_\alpha}\CP_\alpha~\delta(\bk+\bk_\alpha)~\right]
\nonumber \\
& & +{1 \over(2\pi)^3}{2\mu_\gamma \over \hbar^2}
{1 \over k_\gamma^2-k^2+i \varepsilon}
~2~T_{\gamma \alpha}(\bk, \bk_\alpha; E)\ \ .
\label{fm19}
\end{eqnarray}
If we use \eq{fm19} in \eq{fm18} and release
the on-shell condition \eq{fm8}, we finally obtain
\begin{eqnarray}
& & T_{\gamma \alpha}(\bp, \bq; E)
=V_{\gamma \alpha}(\bp, \bq; E)
+\sum_\beta {1 \over (2\pi)^3} \int d \bk
~V_{\gamma \beta}(\bp, \bk; E) \nonumber \\
& & \times {2 \mu_\beta \over \hbar^2}
{1 \over k_\beta^2-k^2+i \varepsilon}
~T_{\beta \alpha}(\bk, \bq; E)\ \ ,
\label{fm20}
\end{eqnarray}
which we call an LS-RGM equation.

The partial-wave decomposition of the scattering amplitudes etc. are
carried out in the standard way \cite{HT70}. For example, we have
\footnote{In this paper we use the phase convention
of the $S$-matrix defined through
the partial wave decomposition in terms of the
time reversal state $i^\ell Y_{\ell m}(\hat{\br})$ in the coordinate
space. This phase convention is different from the standard one
by Blatt and Biedenharn \cite{BL52}
by the factor $i^{\ell-\ell^\prime}$, and introduces an extra minus
sign for the mixing parameter $\epsilon_J$ for the $NN$ scattering.
Similarly, all the Pauli-spinor invariants in \eq{in1} are
taken to be ``real'' operators \protect\cite{FU97}.}
\begin{eqnarray}
& & T_{\gamma \alpha}(\bk, \bk^\prime; E)
=\sum_{JM\ell \ell^\prime}^{\qquad \prime}
4\pi~T^J_{\gamma S^\prime \ell^\prime,\alpha S \ell}(k, k^\prime; E)
\nonumber \\
& & \times \sum_{m^\prime} \langle \ell^\prime m^\prime
S^\prime S_z^\prime | JM \rangle
~Y_{\ell^\prime m^\prime}(\widehat{\bk})
~\sum_{m} \langle \ell m S S_z | JM \rangle
~Y_{\ell m}^*(\widehat{\bk}^\prime)\ \ ,
\label{fm21}
\end{eqnarray}
where $\gamma=\left[1/2(11)\,c_1, 1/2(11)c_2 \right]$
$S^\prime S^\prime_z YII_z; \CP^\prime$.
The prime on the summation symbol indicates 
that $\ell$ ($\ell^\prime$) is limited to such values that satisfy
the condition $(-1)^\ell=(-1)^S\CP$ (\,$(-1)^{\ell^\prime}
=(-1)^{S^\prime}\CP^\prime$\,).
We write the spin quantum
numbers $S$ and $S^\prime$ of the $T-$matrix 
explicitly for the later convenience,
although these are already included in the definition
of $\alpha$ and $\gamma$.
The partial-wave decomposition of the LS-RGM equation
yields
\begin{eqnarray}
& & T^J_{\gamma S^\prime \ell^\prime, \alpha S \ell}(p, q; E)
=V^J_{\gamma S^\prime \ell^\prime, \alpha S \ell}(p, q; E)
+\sum_{\beta S^{\prime \prime} \ell^{\prime \prime}}^{\qquad \prime}
{4\pi \over (2\pi)^3} \int^{\infty}_{0} k^2\,d\,k \nonumber \\
& & \times V^{J}_{\gamma S^\prime \ell^\prime, \beta S^{\prime \prime}
\ell^{\prime \prime}}(p, k; E)
~{2 \mu_\beta \over \hbar^2}
{1 \over k_\beta^2-k^2+i \varepsilon}
~T^J_{\beta S^{\prime \prime} \ell^{\prime \prime}, \alpha S \ell}
(k, q; E)\ \ ,
\label{fm22}
\end{eqnarray}
where $E=E_\beta^{int}+(\hbar^2/2 \mu_\beta)
\bk_\beta^2$ and $V^J_{\gamma S^\prime \ell^\prime,
\alpha S \ell}(p, q; E)$ is the partial-wave decomposition
of $V_{\gamma \alpha}(\bp, \bq; E)$.
The Lippmann-Schwinger equation \eq{fm22} involves a pole
at $k=k_\beta$ in the Green function.
A proper treatment of such a singularity for positive energies
is well known in the field of few-body
problems. \cite{NO65,KO65,BJ76} 
Here we use the technique developed by
Noyes \cite{NO65} and Kowalski \cite{KO65},
and separate the momentum region of $k$ (and also $p$ and $q$) into
two pieces. After eliminating the singularity,
we carry out the integral over $0 \leq k \leq k_\beta$ by
the Gauss-Legendre 15-point quadrature formula,
and the integral over $k_\beta \leq k < \infty$ by
using the Gauss-Legendre 20-point quadrature formula
through the mapping, $k=k_\beta+\tan (\pi/4)(1+x)$.

\subsection{Basic Born kernel}

In this subsection we explain how to obtain the basic Born kernel, 
\eq{fm3}, in detail.
In fact the calculation of the kernel is rather tedious
and a careful check of the calculation must be made. 
Recent modern techniques, especially
developed in the microscopic nuclear cluster theory
of light nuclei \cite{SUP77,SU83,KI94}, have greatly reduced
the labor of tedious calculations. 
Since two-body forces used in such applications
are usually momentum-independent central and spin-orbit forces
of the Gaussian radial dependence,
one needs to extend the technique to more general two-body forces
involving various types of the tensor forces
and Yukawa functions \cite{FU86,FU87}. Appendix A gives a very
general formula to calculate the Born kernel directly from
such two-body forces by simple Gaussian integrations.

Another important technical development in our quark-model
approach to the $NN$ and $YN$ interactions is motivated by 
the rich flavor contents of the spin-flavor $SU_6$ wave functions 
of baryons. The operator formalism introduced
in Refs.\,\citen{FU87s} and \citen{FU92} makes it possible to represent
this flavor dependence in a transparent form by using
abstract $SU_3$ operators expressed by the basic electric- and
magnetic-type $SU_6$ unit vectors.
This formalism is also useful to deal with the flavor symmetry
breaking.\footnote{The full account of the operator formalism
for the spin-flavor factors of the RGM kernel will be published
elsewhere.} 

Keeping in mind the operator formalism of spin-flavor factors,
we can express the basic Born kernel in \eq{fm3} as
\begin{eqnarray}
M^B(\bq_f, \bq_i; E)
& = & \langle\,e^{i \bq_f \cdot \bR}\,\vert
\,G(\bR, \bR^\prime; E)
\,\vert\,e^{i \bq_i \cdot {\bR}^\prime} \rangle \nonumber \\
& = &M_D^{CN}(\bq_f, \bq_i)
+M_D^{SN}(\bq_f, \bq_i)
+M_D^{TN}(\bq_f, \bq_i)\,S_{12}(\bk,\bk) \nonumber \\
& & +\sum_\Omega M^\Omega(\bq_f, \bq_i)\,\CO^\Omega(\bq_f, \bq_i)
-\varepsilon~M_N(\bq_f, \bq_i)\ \ ,
\label{fm23}
\end{eqnarray}
where $\varepsilon$ is the relative energy
in the final channel (in the prior form) when
the channel matrix elements are taken at the baryon level. 
Each component of the Born kernel \eq{fm23} is given in Appendix B
in terms of the transferred momentum $\bk=\bq_f-\bq_i$ and
the local momentum $\bq=(\bq_f+\bq_i)/2$.
In \eq{fm23} the space-spin
invariants $\CO^\Omega=\CO^\Omega(\bq_f, \bq_i)$ are
given by $\CO^{central}=1$ and\footnote{Here we use a slightly
different notation for $\Omega$ from that used in $\S$\,2.1,
but the correspondence is almost apparent.}
\begin{eqnarray}
& & \CO^{LS} = i \bn \cdot \bS\ ,\quad
\CO^{LS^{(-)}} = i \bn \cdot \bS^{(-)}\ ,\quad 
\CO^{LS^{(-)}\sigma} = i \bn \cdot \bS^{(-)}\,P_\sigma\ \ ,
\nonumber \\
& & \hbox{with} \quad \bn=[ \bq_i \times \bq_f ]\ ,\quad 
\bS={1 \over 2}(\bfsigma_1+\bfsigma_2)\ ,\quad
\bS^{(-)}={1 \over 2}(\bfsigma_1-\bfsigma_2)\ ,\nonumber \\
& & \hbox{and} \quad P_\sigma={1+\bfsigma_1
\cdot \bfsigma_2 \over 2}\ \ .
\label{fm24}
\end{eqnarray} 
For the tensor part, it would be convenient to take three
natural operators defined by
\begin{equation}
\CO^{T} = S_{12}(\bk, \bk)\ ,\quad
\CO^{T^\prime} = S_{12}(\bq, \bq)\ ,\quad 
\CO^{T^{\prime \prime}} = S_{12}(\bk, \bq)\ \ ,
\label{fm26}
\end{equation}
where $S_{12}(\ba, \bb)=(3/2) [\,({\bfsigma}_1 \cdot \ba)
(\bfsigma_2 \cdot \bb)+(\bfsigma_2 \cdot \ba)
(\bfsigma_1 \cdot \bb)\,]
-(\bfsigma_1 \cdot \bfsigma_2)(\ba \cdot \bb)$.
The invariant Born kernel $M^\Omega (\bq_f, \bq_i)$ in
\eq{fm23} consists of various types of spin-flavor
factors $X_{\CT}^\Omega$ and
the spatial functions $f_{\CT}^\Omega (\theta)$ calculated 
for the quark-exchange kernel
of the FB interaction.\footnote{Here we will show only the expressions
for the quark sector. Those for the EMEP sector are easily obtained
by some trivial modifications.}
It is explicitly given in Appendix B and is generally expressed as
\begin{equation}
M^\Omega (\bq_f, \bq_i)=\sum_\CT X_\CT^\Omega~f_\CT^\Omega
(\theta)\ \ .
\label{fm27}
\end{equation}
Here we should note that the spin-flavor factors depend
on the isospin and that the spatial
function $f_{\CT}^\Omega (\theta)$ is
actually a function of $\bq_f^2$, $\bq_i^2$ and 
the relative angle $\theta$: $\cos \theta=\widehat{\bq}_f
\cdot \widehat{\bq}_i$. The sum over $\CT$ in \eq{fm27} is
with respect to the quark-exchange interaction
types \cite{KI94} $\CT=E$, $S$, $S^\prime$, $D_+$ and $D_-$.
The factors $X_E^\Omega$ (possible only for the central
force) should be replaced
with $-X_{S^\prime}^\Omega$, because of the subtraction
of the internal-energy contribution in the prior form.
Finally the partial-wave decomposition of the Born kernel,
$V^{J\Omega}_{S^\prime \ell^\prime, S \ell}(q_f, q_i)$ can be
calculated by using\footnote{We use the Gauss-Legendre
20-point quadrature formula to carry out the numerical integration
of \protect\eq{fm28} over $\cos \theta=-1 \sim 1$.}  
\begin{equation}
f^\Omega_{\CT \ell}={1 \over 2} \int_{-1}^1 
f^\Omega_{\CT}(\theta)\,P_\ell(\cos \theta)~d (\cos \theta)\ \ ,
\label{fm28}
\end{equation}
as follows:
\begin{eqnarray}
& & V^{J~central~\Omega}_{S^\prime \ell^\prime, S \ell}(q_f, q_i)
=\delta_{\ell^\prime, \ell} \delta_{S^\prime, S}
\sum_\CT X_\CT^\Omega~f_{\CT \ell}^\Omega
\quad \hbox{with} \quad (\bfsigma_1 \cdot \bfsigma_2)
=2S(S+1)-3\ \ , \nonumber \\ 
& & V^{J~LS}_{S^\prime \ell^\prime, S \ell}(q_f, q_i)
=\delta_{\ell^\prime, \ell} \delta_{S^\prime, S} \delta_{S, 1}
~q_f q_i {1 \over 2(2\ell+1)} \left[ \ell(\ell+1)+2-J(J+1) \right]
\nonumber \\
& & \ \ \hspace{25mm} \times \sum_\CT X_\CT^{LS}~(f_{\CT \ell+1}^{LS}
-f_{\CT \ell-1}^{LS})\ \ , \nonumber \\
& & \left.
\begin{array}{c}
V^{J~LS^{(-)}}_{S^\prime \ell^\prime, S \ell}(q_f, q_i) \\
V^{J~LS^{(-)}\sigma}_{S^\prime \ell^\prime, S \ell}(q_f, q_i) \\
\end{array} \right\}
=\left\{ \begin{array}{c}
-1 \\
(-1)^S \\
\end{array} \right\}
\delta_{\ell^\prime, \ell}\delta_{J, \ell}
~q_f q_i {\sqrt{J(J+1)} \over 2J+1} \nonumber \\
& & \ \ \hspace{25mm} \times
\sum_\CT \left\{
\begin{array}{c}
X_\CT^{LS^{(-)}} \\
X_\CT^{LS^{(-)}\sigma} \\
\end{array} \right\}
~(f_{\CT J+1}^{LS}
-f_{\CT J-1}^{LS})
\quad \hbox{for} \quad S,~S=1,~0~\hbox{or}~0,~1\ \ , \nonumber \\
& & V^{J~T~total}_{S^\prime \ell^\prime, S \ell}(q_f, q_i)
=\delta_{S^\prime, S} \delta_{S, 1}~(S_{12})^J_{\ell^\prime, \ell}
\nonumber \\
& & \times \left[\,q_f^2 V_\ell^{T(ff)}
+q_i^2 V_{\ell^\prime}^{T(ii)}+q_f q_i
\left\{
\begin{array}{c}
V_J^{T(fi)} \\
{1 \over 2\ell+1}\left[ \left(\ell-{1 \over 2}\right) V_{\ell+1}^{T(fi)}
+\left(\ell+{3 \over 2}\right) V_{\ell-1}^{T(fi)} \right] \\
\end{array} \right\}~\right] \nonumber \\
& & \ \ \hspace{25mm} \hbox{for} \quad \left\{
\begin{array}{c}
\ell^\prime=\ell \pm 2 \quad \hbox{and} \quad J=\ell \pm 1 \\
\ell=\ell^\prime=J,~J \pm 1 \\
\end{array} \right.\ \ .
\label{fm29}
\end{eqnarray} 
Here the tensor component is the sum over $\Omega=T$,
$T^\prime$ and $T^{\prime \prime}$ and 
\begin{eqnarray}
\left.
\begin{array}{c}
V_\ell^{T(ff)} \\
V_\ell^{T(ii)} \\
\end{array} \right\}
& = & \left\{
\begin{array}{c}
9 \\ 1 \\
\end{array} \right\}
{1 \over 4} X_S^T~f_{S \ell}^T
+\left\{
\begin{array}{c}
1 \\ 9 \\
\end{array} \right\}
{1 \over 4} X_{S^\prime}^T~f_{S^\prime \ell}^T
+X_{D+}^T~f_{D+ \ell}^T+{1 \over 4} X_{D-}^T~f_{D- \ell}^T
\ \ ,\nonumber \\
V_\ell^{T(fi)} & = & -{3 \over 2} \left[ X_S^T~f_{S \ell}^T
+ X_{S^\prime}^T~f_{S^\prime \ell}^T \right]
-2 X_{D+}^T~f_{D+ \ell}^T+{1 \over 2} X_{D-}^T~f_{D- \ell}^T\ \ .
\label{fm30}
\end{eqnarray} 
Furthermore, $(S_{12})^J_{\ell^\prime, \ell}$ is the standard
tensor factor. In Eqs.\,(\ref{fm27}), (\ref{fm29}) and (\ref{fm30}),
the spin-flavor factors $X_\CT^\Omega$ should be replaced
with $X_{x=1 \CT}^\Omega$ for the EMEP exchange 
terms, $\Omega=CN,~SN$ and $TN$. Also, the direct term with $x=0$
is possible only for $\CT=D_+$.

\subsection{Invariant amplitudes}

Since the three vectors, $\bk=\bq_f-\bq_i$, $\bq=(\bq_f+\bq_i$)/2 and
$\bn=\bq_i \times \bq_f=\bq \times \bk$ are not
mutually orthogonal for the $YN$ interaction,
we use $\bk=\bq_f-\bq_i$, $\bn=\bq \times \bk$ and
$\bP=\bk \times \bn=\bk^2 \bq-(\bk \cdot \bq)\bk$ to define
the invariant amplitudes of \eq{fm14}:   
\begin{eqnarray}
& & M(\bq_f, \bq_i; E)=g_0+h_0~i(\bfsigma_1+\bfsigma_2)
\cdot \widehat{\bn}
+h_-~i(\bfsigma_1-\bfsigma_2)\cdot \widehat{\bn}
\nonumber \\
& & +h_n~(\bfsigma_1\cdot \widehat{\bn})
\cdot (\bfsigma_2\cdot \widehat{\bn})
+h_k~(\bfsigma_1\cdot \widehat{\bk})
\cdot (\bfsigma_2\cdot \widehat{\bk})
+h_P~(\bfsigma_1\cdot \widehat{\bP})
\cdot (\bfsigma_2\cdot \widehat{\bP})
\nonumber \\
& & +f_+ \left\{(\bfsigma_1\cdot \widehat{\bk})
(\bfsigma_2\cdot \widehat{\bP})+(\bfsigma_1\cdot \widehat{\bP})
(\bfsigma_2\cdot \widehat{\bk})\right\} \nonumber  \\
& & +f_- \left\{(\bfsigma_1\cdot \widehat{\bk})
(\bfsigma_2\cdot \widehat{\bP})-(\bfsigma_1\cdot \widehat{\bP})
(\bfsigma_2\cdot \widehat{\bk})\right\}\ \ .
\label{in1}
\end{eqnarray}
The eight invariant amplitudes, $g_0, \cdots, f_-$, are complex
functions of the total energy $E$ and the scattering
angle, $\cos \theta=\widehat{\bq}_i \cdot \widehat{\bq}_f$.
Three of the eight invariant amplitudes,
$h_-$, $f_+$ and $f_-$, do not appear for the $NN$ scattering
due to the identity of two particles ($h_-$ and $f_-$) and
the time-reversal invariance ($f_-$ and $f_+$).
These three terms correspond to
the non-central forces characteristic in the $YN$ scattering;
i.e., $h_-$ corresponds to $LS^{(-)}$,
$f_+$ to $S_{12}(\br, \bp)$, and $f_-$ to $LS^{(-)}\sigma$,
respectively. \cite{FU97}
In particular, the antisymmetric $LS$ interactions,
$LS^{(-)}$ and $LS^{(-)}\sigma$,
involve the spin change between 0 and 1, together with
the transition of the flavor-exchange symmetry $\CP \neq \CP'$.
In the $NN$ scattering this process is not allowed,
since the flavor-exchange symmetry is uniquely specified
by the conserved isospin: $\CP=(-1)^{1-I}$.
On the contrary, these interactions are in general all possible
in the $YN$ scattering, which gives an intriguing interplay
of non-central forces.

The invariant amplitudes are expressed by the $S$-matrix elements
through the partial-wave decomposition of 
the scattering amplitude \eq{fm14}.
If we write it as
\begin{eqnarray}
& & M_{\gamma \alpha}(\bq_f, \bq_i; E)
=\sum_{JM\ell \ell^\prime}^{\qquad \prime}
4\pi~R^J_{\gamma S^\prime \ell^\prime,\alpha S \ell}
\nonumber \\
& & \times \sum_{m^\prime} \langle \ell^\prime m^\prime
S^\prime S_z^\prime | JM \rangle
~Y_{\ell^\prime m^\prime}(\widehat{\bq}_f)
~\sum_{m} \langle \ell m S S_z | JM \rangle
~Y_{\ell m}^*(\widehat{\bq}_i)\ \ ,
\label{in2}
\end{eqnarray}
the partial-wave component $R^J_{\gamma S^\prime \ell^\prime,
\alpha S \ell}=(1/2i)(S^J_{\gamma S^\prime \ell^\prime,
\alpha S \ell}-\delta_{\gamma, \alpha} \delta_{S^\prime, S}
\delta_{\ell^\prime, \ell})$ is obtained
from the relationship in \eq{fm17}:
\begin{equation}
R^J_{\gamma S^\prime \ell^\prime, \alpha S \ell}
=-{\sqrt{\mu_\gamma \mu_\alpha q_f q_i} \over 2\pi \hbar^2}
~T^J_{\gamma S^\prime \ell^\prime, \alpha S \ell}(q_f, q_i; E)
\label{in3}
\end{equation}
where $\varepsilon_\gamma=(\hbar^2/2\mu_\gamma)q_f^2$, 
$\varepsilon_\alpha=(\hbar^2/2\mu_\alpha)q_i^2$ 
and $E=E^{int}_\gamma+\varepsilon_\gamma
=E^{int}_\alpha+\varepsilon_\alpha$.
The formulae given in Appendix D are then used to reconstruct
the invariant amplitudes by the solution
of the LS-RGM equation \eq{fm22}.
All the scattering observables are expressed
in terms of these invariant amplitudes.
For example, the differential cross section and the
polarization of the scattered particle are
given by
\begin{eqnarray}
{d \sigma \over d \Omega} & = & \sigma_0(\theta)
= |g_0|^2 + |h_0+h_-|^2+|h_0-h_-|^2\ \ , \nonumber \\
& & +|h_n|^2+|h_k|^2+|h_P|^2
+|f_++f_-|^2+|f_+-f_-|^2 \nonumber \\
P(\theta) & = & 2 \Im m \left[ g_0(h_0+h_-)^*+
h_n (h_0-h_-)^*\right]
\nonumber \\
& & +2 \Im m \left[ h_k(f_+-f_-)^*-h_P(f_++f_-)^* \right]\ \ .
\label{in4}
\end{eqnarray}
\begin{figure}[t]
\begin{center}
\begin{minipage}[t]{\textwidth}
\epsfxsize=\textwidth
\centerline{\epsfbox{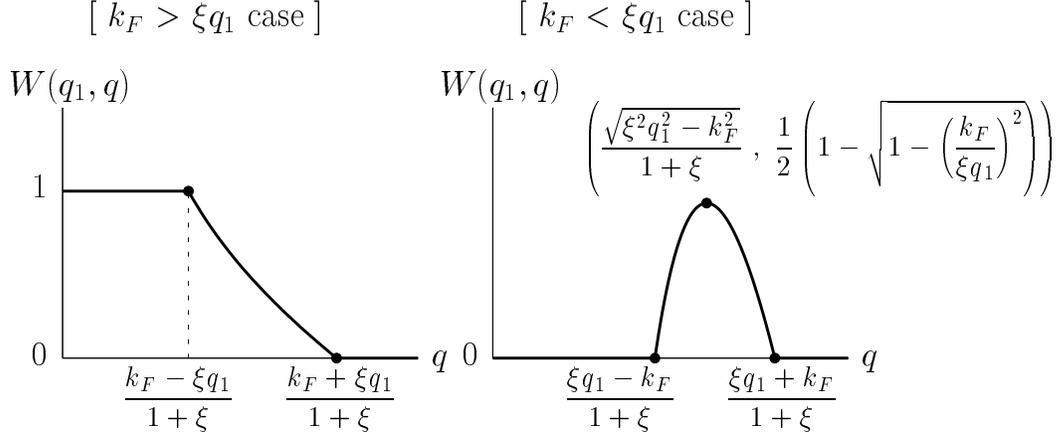}}
\vspace{0mm}
\caption{The weight function $W(q_1, q)$ in
\protect\eq{sp5} as a function of $q$.}
\label{fig1}
\end{minipage}
\end{center}
\end{figure}

\subsection{$t^{eff}\rho$ prescription
for single-particle potentials}

In this subsection we will derive an approximate formula
to calculate s.p. potentials appearing in the
lowest-order Brueckner theory. 
The Bethe-Goldstone (BG) equation for the $G$-matrix solution
is given by \cite{KO99}
\begin{eqnarray}
& & G^J_{\gamma S^\prime \ell^\prime, \alpha S \ell}(p, q; K, \omega)
=V^J_{\gamma S^\prime \ell^\prime, \alpha S \ell}(p, q; E)
+\sum_{\beta S^{\prime \prime} \ell^{\prime \prime}}^{\qquad \prime}
{4\pi \over (2\pi)^3} \int^{\infty}_{0} k^2\,d\,k \nonumber \\
& & \times V^J_{\gamma S^\prime \ell^\prime, \beta S^{\prime \prime}
\ell^{\prime \prime}}(p, k; E)
~{Q_\beta(k, K) \over e_\beta(k, K; \omega)}
~G^J_{\beta S^{\prime \prime} \ell^{\prime \prime}, \alpha S \ell}
(k, q; K, \omega)\ \ .
\label{sp1}
\end{eqnarray}
where we assume $E=E_\alpha^{int}+(\hbar^2/2 \mu_\alpha) q^2$.
In \eq{sp1}, $Q_\beta(k, K)$ stands for
the angle-averaged Pauli operator
and $e_\beta(k, K; \omega)$ is the energy denominator
given by\footnote{For s.p. potentials,
we use the notation $b_1=b$ and $a_1=a$ to specify baryons.}
\begin{equation}
e_\beta(k, K; \omega)=\omega-E_{b}(k_1)-E_N(k_2)\ \ ,
\label{sp2}
\end{equation}
with the s.p. energy $E_b(k)$:
\begin{equation}
E_b(k)=M_b+{\hbar^2 \over 2M_b} k^2+U_b(k)\ \ .
\label{sp3}
\end{equation}
The s.p. potential $U_b(k)$ is calculated from
\begin{eqnarray}
& & U_a(q_1)=(1+\delta_{a,N})(1+\xi)^3 \sum_I
{2I+1 \over 2(2I_a+1)} \nonumber \\
& & \times \sum_{J\ell S} (2J+1) {1 \over 2 \pi^2}
\int_{0}^{q_{max}} q^2\,d\,q
~W(q_1, q)~G_{aS\ell, aS\ell}^J (q, q; K, \omega)\ \ ,
\label{sp4}
\end{eqnarray}
where $q_{max}=(k_F+\xi q_1)/(1+\xi)$ with $\xi=(M_N/M_a)$ and $W(q_1,
q)$ is the phase space factor given by 
\begin{equation}
W(q_1, q)={1 \over 2}\left(\,1-[-1|x_0|1]\,\right)\ ,
\quad x_0={\xi^2 q_1^2+(1+\xi)^2 q^2
-k_F^2 \over 2 \xi (1+\xi) q_1 q}\ ,
\label{sp5}
\end{equation}
with $[a|b|c] \equiv \max (a, \min (b,c))$ \cite{SCHU}.
The profile of $W(q_1, q)$ is illustrated in Fig.\,1.
The starting energy $\omega$ in \eq{sp2} is a sum
of the s.p. energies of two interacting baryons:
\begin{eqnarray}
& & \omega = E_a(q_1)+E_N(q_2) \nonumber \\
& & = M_a+M_N+{\hbar^2 \over 2(M_a+M_N)} K^2
+{\hbar^2 \over 2 \mu_\alpha} q^2
+U_a(q_1)+U_N(q_2)\ \ ,
\label{sp6}
\end{eqnarray}
where $\bK$ and $\bq$ are the total and relative momenta
corresponding to the initial s.p. momenta $\bq_1$ and $\bq_2$.
Once $q_1$ and $q$ are given,
the values of $K$ and $\omega$ in \eq{sp4} are
calculated through
\begin{eqnarray}
& & K=(1+\xi) \left[q_1^2+q^2-q_1 q \left( 1+[-1|x_0|1] \right)
~\right]^{{1 \over 2}}\ \ , \nonumber \\
& & q_2=\left[ {\xi \over 1+\xi} K^2+(1+\xi) q^2 - \xi q_1^2
~\right]^{{1 \over 2}}\ \ ,\nonumber \\
& & \omega=E_a(q_1)+E_N(q_2)\ \ . 
\label{sp7}
\end{eqnarray}

Let us specialize the system to $NN$ and consider the
asymptotic behavior of the s.p. potential $U_N(q_1)$ in symmetric 
nuclear matter. 
When $q_1$ is sufficiently large,
the weight function $W(q_1, q)$ shows a delta-function like
behavior around $q=q_1/2$ with respect
to the variable $q$. (See Fig.\,1.)
We can replace the $G$-matrix in \eq{sp4} with the one 
evaluated at $q=q_1/2$, and carry out the integration over $q$ by
\begin{eqnarray}
& & \int_{0}^{q_{max}} q^2\,d\,q
~W(q_1, q) = {k_F^3 \over 3}{1 \over (1+\xi)^3}\ \ .
\label{sp8}
\end{eqnarray}
On the other hand, the $G$-matrix equation \eq{sp1}
for $q \rightarrow \infty$ should approach the LS-RGM
equation \eq{fm22}, since the Pauli operator $Q_\beta(k, K)$ plays
a minor role in such a high momentum region,
and the s.p. potentials in the energy
denominator $e_\beta(k, K; \omega)$ is relatively unimportant
in comparison with the large kinetic energies.
Thus we can expect that the s.p. potential in the high momentum
region is well approximated by a product
of the on-shell $T$-matrix with $q=q_1/2$ and the
density of nuclear matter, which is related to the Fermi
momentum $k_F$ with $\rho=(2/3\pi^2)\,k^3_F$.
It should be noted that only the spin-isospin independent
part of the $T$-matrix contributes, since we take spin-isospin
sum for the target nucleons. If we note the relationship \eq{in3}
and use the partial-wave decomposition
of the invariant amplitude $g_0$ given in \eq{d2},
we can easily show
\begin{equation}
U_N(q_1) = {1 \over 4} U^0_N(q_1)+{3 \over 4} U^1_N(q_1)
\quad \hbox{with} \quad U^I_N(q_1) = 2\rho~g^I_0(q=q_1/2, \theta=0)
\ \ , 
\label{sp9}
\end{equation}
except for the overall factor $-(2 \pi \hbar^2/\mu q)$.
This relationship is also shown in terms
of $np$ and $pp$ invariant amplitudes:
\begin{eqnarray}
U_N(q_1) & = & (\rho/2) \left(g^0_0(q_1/2, 0)+g^1_0(q_1/2, 0)\right)
+(\rho/2) 2g^1_0(q_1/2, 0) \nonumber \\
& = & (\rho/2) (g^{np}_0(q_1/2, 0)+g^{pp}_0(q_1/2, 0))\ \ ,
\label{sp10}
\end{eqnarray}

\begin{wraptable}{c}{\textwidth}
\caption{Comparison of the phase-shift values
and mixing angles (in degree),
calculated by the LS-RGM formalism (lseq) and the improved variational
method (var), for the $\Lambda N$
$\hbox{}^3S_1$-$\hbox{}^3D_1$ - $\Sigma N$
$\hbox{}^3S_1$-$\hbox{}^3D_1$ (upper)
and $\Lambda N$ $\hbox{}^1P_1$-$\hbox{}^3P_1$ - $\Sigma N$
$\hbox{}^1P_1$-$\hbox{}^3P_1$ (lower) coupled-channel problems.
The mixing angles, $\epsilon_1$ and $\rho_1$,
are only approximately defined for the energies above
the $\Sigma N$ threshold.
The phase convention of $\epsilon_1$ is the standard one
for the bar phase shifts. The model is RGM-F \protect\cite{FU95}.
}
\label{table1}
\bigskip
\begin{center}
\renewcommand{\arraystretch}{1.3}
\setlength{\tabcolsep}{3mm}
\begin{tabular}{rrrrrrr}
\hline
\hline
$p_{\Lambda}$ & & \multicolumn{2}{c}{$\Lambda N$}
 & \multicolumn{2}{c}{$\Sigma N$} & $\Lambda N$ \\
\cline{3-7}
 & & $^3S_1$ & $^3D_1$ & $^3S_1$ & $^3D_1$
& $\epsilon_1$ \\
\hline
    & lseq & 37.494 & 8.656 & $-$ & $-$ & 13.802 \\
430 & var  & 37.559 & 8.654 & $-$ & $-$ & 13.814 \\
\cline{2-7}
    & diff  & $-0.065$ & 0.002 & $-$ & $-$ & $-0.012$ \\
\hline
    & lseq & 49.169 & 15.227 & $-$ & $-$ & 22.999 \\
440 & var  & 49.263 & 15.212 & $-$ & $-$ & 23.016 \\
\cline{2-7}
    & diff & $-0.094$ & 0.017 & $-$ & $-$ & $-0.017$ \\
\hline
    & lseq & 12.831 & $-5.535$ & 122.619 & $-0.014$ & $-21.090$ \\
450 & var  & 12.825 & $-5.528$ & 122.708 & $-0.011$ & $-21.099$ \\
\cline{2-7}
    & diff & 0.006 & $-0.007$ & $-0.089$ & $-0.003$ & 0.009 \\
\hline
    & lseq & 14.793 & $-4.461$ & 87.265 & $-0.165$ & $-13.440$ \\
460 & var  & 14.801 & $-4.458$ & 87.298 & $-0.152$ & $-13.449$ \\
\cline{2-7}
    & diff & $-0.008$ & $-0.003$ & $-0.033$ & $-0.013$ & 0.009 \\
\hline
\end{tabular}
\end{center}
\vspace{5mm}
\begin{center}
\renewcommand{\arraystretch}{1.3}
\setlength{\tabcolsep}{3mm}
\begin{tabular}{rrrrrrr}
\hline
\hline
$p_{\Lambda}$ & & \multicolumn{2}{c}{$\Lambda N$}
 & \multicolumn{2}{c}{$\Sigma N$} & $\Lambda N$ \\
\cline{3-7}
 & & $^1P_1$ & $^3P_1$ & $^1P_1$ & $^3P_1$
& $\rho_1$ \\
\hline
    & lseq & 9.161 & 5.195 & $-$ & $-$ & $-8.213$ \\
300 & var  & 9.161 & 5.188 & $-$ & $-$ & $-8.208$ \\ 
\cline{2-7}
    & diff & 0.000 & 0.007 &  &  & $-0.005$ \\
\hline
    & lseq & 82.874 & 0.757 & $-$ & $-$ & $-41.647$ \\
400 & var  & 82.605 & 0.971 & $-$ & $-$ & $-41.636$ \\
\cline{2-7}
    & diff & 0.269 & $-0.214$ & $-$ & $-$ & $-0.011$ \\
\hline
    & lseq & $-15.697$ & $-20.945$ & $-1.812$ & $-8.113$ & 16.559 \\
500 & var  & $-15.690$ & $-20.944$ & $-1.812$ & $-8.120$ & 16.563 \\
\cline{2-7}
    & diff & $-0.007$ & $-0.001$  & 0.000    & 0.007    & $-0.004$ \\
\hline
    & lseq & $-14.077$ & $-21.357$ & $-5.725$ & $-22.826$ & 13.399 \\
600 & var  & $-14.070$ & $-21.358$ & $-5.724$ & $-22.831$ & 13.401 \\
\cline{2-7}
    & diff & $-0.007$ & 0.001 & $-0.001$ & 0.005 & $-0.002$ \\
\hline
\end{tabular}
\end{center}
\end{wraptable}

\noindent
where $\rho/2$ is the proton and neutron density.
This implies that the asymptotic behavior of the s.p. potential
in the high momentum region is determined by the values
of the spin-independent central invariant amplitudes
at the forward angle. This rule is called $t^{eff} \rho$ prescription,
which is sometimes used in the relativistic mean field theory.
For example, if we want to
calculate $U_N(q_1)$ for $q_1=6.210\hbox{fm}^{-1}$ in
nuclear matter of the normal density
with $k_F=1.35~\hbox{fm}^{-1}$ ($\rho=(2/3\pi^2)\,k^3_F
=0.1662~\hbox{fm}^{-3}$),
we only need to derive the invariant amplitude
$(g^0_0(\theta=0)+3g^1_0(\theta=0))$ for $T_{lab}=800$ MeV
($q=q_1/2=3.105~\hbox{fm}^{-1}$) and multiply it by
the factor, $-(4\pi/q)(\hbar^2/M_N)(\rho/2)=-(4\pi/3.105)
\cdot 41.47 \cdot 0.0831=-13.95$.

\section{Comparison with the improved variational method}

As a check of the LS-RGM formalism, we consider the phase-shift
parameters of the $\Lambda N$-$\Sigma N (I=\hbox{1/2})$
coupled-channel system, and compare them with the predictions 
by the improved variational method.
Here we consider two different types
of the $\Lambda N$-$\Sigma N(I=1/2)$ couplings.
The first one is the $\hbox{}^3S_1$-$\hbox{}^3D_1$ channel
coupling by the tensor force, which dominantly comes from
the EMEP, especially, from the one-pion exchange tensor force.
The other is the $\hbox{}^1P_1$ - $\hbox{}^3P_1$ coupling
by the $LS^{(-)}$ force originating from the FB interaction.
Since the quark model usually predicts very strong $LS^{(-)}$ force,
the $\Lambda p$ scattering observables involve
very rich information on the characters of the non-central forces
in the $YN$ interaction.

Table I shows a comparison between the phase-shift values predicted
by the LS-RGM formalism (lseq) and
the improved variational method (var) with respect to
the $\Lambda N$ $\hbox{}^3S_1$-$\hbox{}^3D_1$-$\Sigma N$
$\hbox{}^3S_1$-$\hbox{}^3D_1$ (upper) and $\Lambda N$
$\hbox{}^1P_1$-$\hbox{}^3P_1$-$\Sigma N$
$\hbox{}^1P_1$-$\hbox{}^3P_1$ (lower) coupled-channel problems.
The mixing angle $\epsilon_1$ ($\rho_1$) between
the $\hbox{}^3S_1$ and $\hbox{}^3D_1$ 
($\hbox{}^1P_1$ and $\hbox{}^3P_1$) channels
of $\Lambda N$ is also compared.
The model is RGM-F\cite{FU95},
which gives the $\Sigma N$ threshold energy
at $p_\Lambda$ = 445 MeV/$c$. In spite of the prominent resonance
behavior in this energy region, the two methods give very similar
values for the phase shifts.
The difference (diff) is less than $0.1^\circ$
for the $\hbox{}^3S_1$-$\hbox{}^3D_1$ coupling.
For the $\hbox{}^1P_1$-$\hbox{}^3P_1$ coupling,
we find that the accuracy deteriorates because of the strong
resonance in the $\Lambda N$ $\hbox{}^1P_1$ channel.
If we avoid this energy region, the accuracy is very good.
The difference is usually less than $0.01^\circ$.

\begin{table}[t]
\caption{
Resonance behavior near the $\Sigma N$ threshold
for the $I=1/2$ states, predicted by RGM-F \protect\cite{FU95},
FSS \protect\cite{FU96a,FU96b} and
RGM-H \protect\cite{FU96b}.
``step'' denotes the step-like resonance and ``disp'' the
dispersion-like resonance.}
\label{table2}
\bigskip
\begin{center}
\renewcommand{\arraystretch}{1.2}
\setlength{\tabcolsep}{3mm}
\begin{tabular}{cccc}
%\begin{tabular}{@{\hspace{0.5cm}}c@{\hspace{0.5cm}}c@{\hspace{0.5cm}}c
%@{\hspace{0.5cm}}c@{\hspace{0.5cm}}}
\hline
\hline
model & RGM-F & FSS & RGM-H \\
\hline
$E_{\Sigma N}^{th}$ (MeV)
& 39 & 77 & 77 \\
$V^C_{\Sigma N(1/2)}(\hbox{}^3S)$ & $-38$ & $-24$ & $-18$ \\
\hline
$\Lambda N $ $\hbox{}^3S_1$ & step & cusp & cusp \\
$\Lambda N $ $\hbox{}^3D_1$ & disp & disp & disp \\
$\Sigma N $ $\hbox{}^3S_1$ & $\delta=180^\circ \downarrow$
& $\delta \lesssim 60^\circ$ & $\delta \lesssim 45^\circ$ \\
$\Sigma N $ $\hbox{}^3D_1$ & $\delta \lesssim 0$ & $\delta \lesssim 0$
& $\delta \lesssim 0$ \\
\hline
$\Lambda N $ $\hbox{}^1P_1$ & step & step & disp \\
$\Lambda N $ $\hbox{}^3P_1$ & disp & disp & disp \\
$\Sigma N $ $\hbox{}^1P_1$ & $\delta \lesssim 0$ & $\delta \lesssim 0$
& $\delta \sim 0 \rightarrow 60^\circ$ \\
$\Sigma N $ $\hbox{}^3P_1$ & $\delta < 0$ & $\delta \lesssim 0$
& $\delta \lesssim 40^\circ$ \\
\hline
\end{tabular}
\end{center}
\end{table}

Since the resonance behavior in the above two coupled-channel systems
is largely model-dependent, we summarize it in Table II for the
three versions of our quark model.  
Here $V^C_{\Sigma N(1/2)}(\hbox{}^3S)$ indicates the strength
of the central attraction in the $\Sigma N (I=1/2)$ channel,
which is evaluated from the $\hbox{}^3S$ effective potential
obtained through the ${\mbf p}=0$ Wigner transform of the
exchange kernel.
We find that the $\Lambda N$ $\hbox{}^3S_1$ resonance in RGM-F
appears as a cusp, when the attraction
of the $\Sigma N(I=1/2)$ channel
is not strongly attractive as in FSS and RGM-H.
Similarly, the $\Sigma N (I=1/2)$ $\hbox{}^3P_1$ resonance
does not move to the $\Lambda N$ $\hbox{}^1P_1$ state in RGM-H,
which has the weakest central attraction
in the $\Sigma N(I=1/2)$ channel among our three models.

%\newpage

\bigskip

\section{$NN$ invariant amplitudes at intermediate energies}

\begin{figure}[t]
\begin{center}
\begin{minipage}[t]{\textwidth}
\epsfxsize=\textwidth
\centerline{\epsfbox{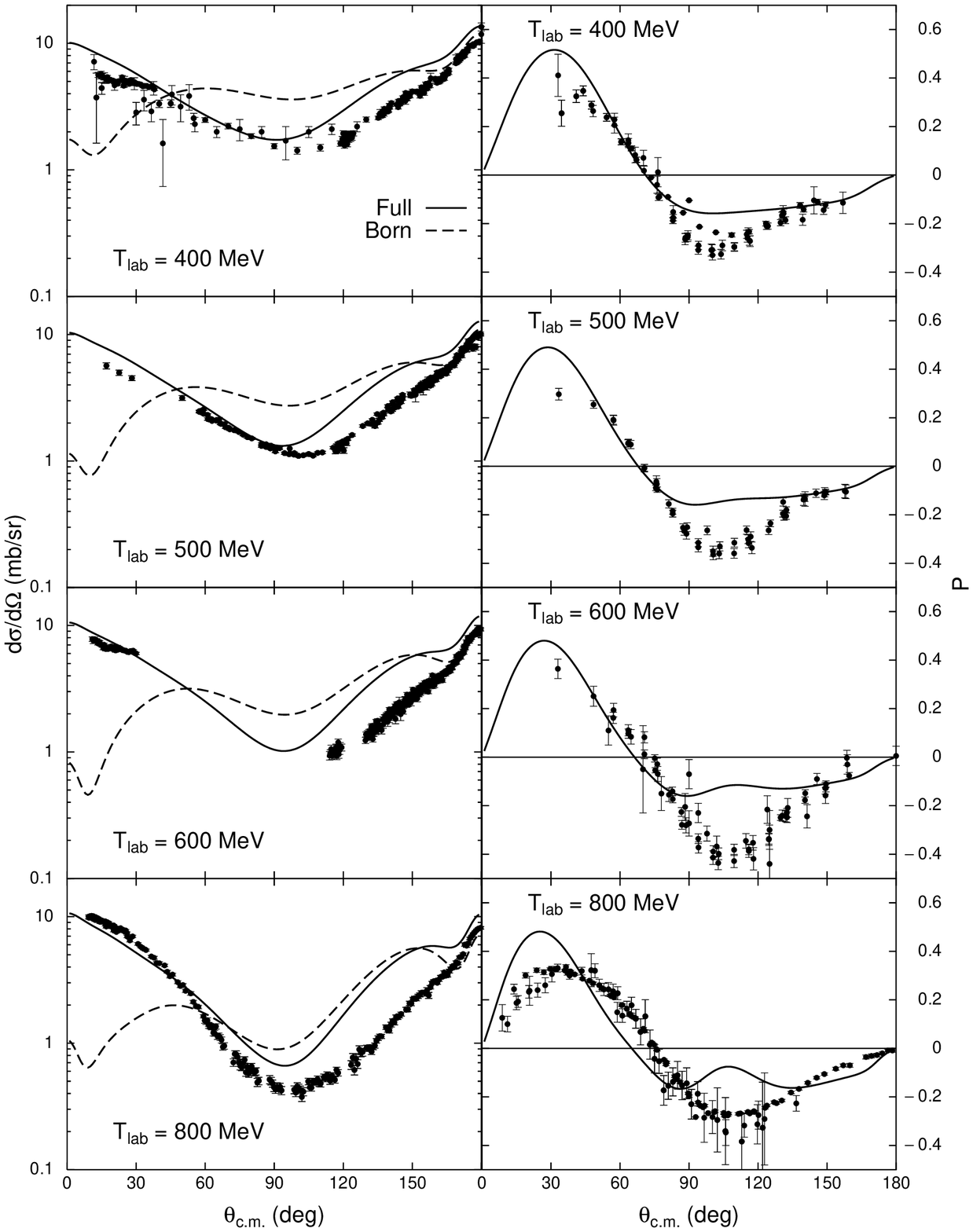}}
\vspace{0mm}
\caption{The differential cross sections and polarization
for the elastic $np$ scattering at $T_{lab} = 400 \sim 800$ MeV.
The model is FSS. The solid curve denotes the full calculation with 
LS-RGM, while the dashed curve the one with the Born approximation.
Experimental data are from \protect\citen{SAID}.}
\label{fig2}
\end{minipage}
\end{center}
\end{figure}
\begin{figure}[b]
\begin{center}
\begin{minipage}[t]{\textwidth}
\epsfxsize=\textwidth
\centerline{\epsfbox{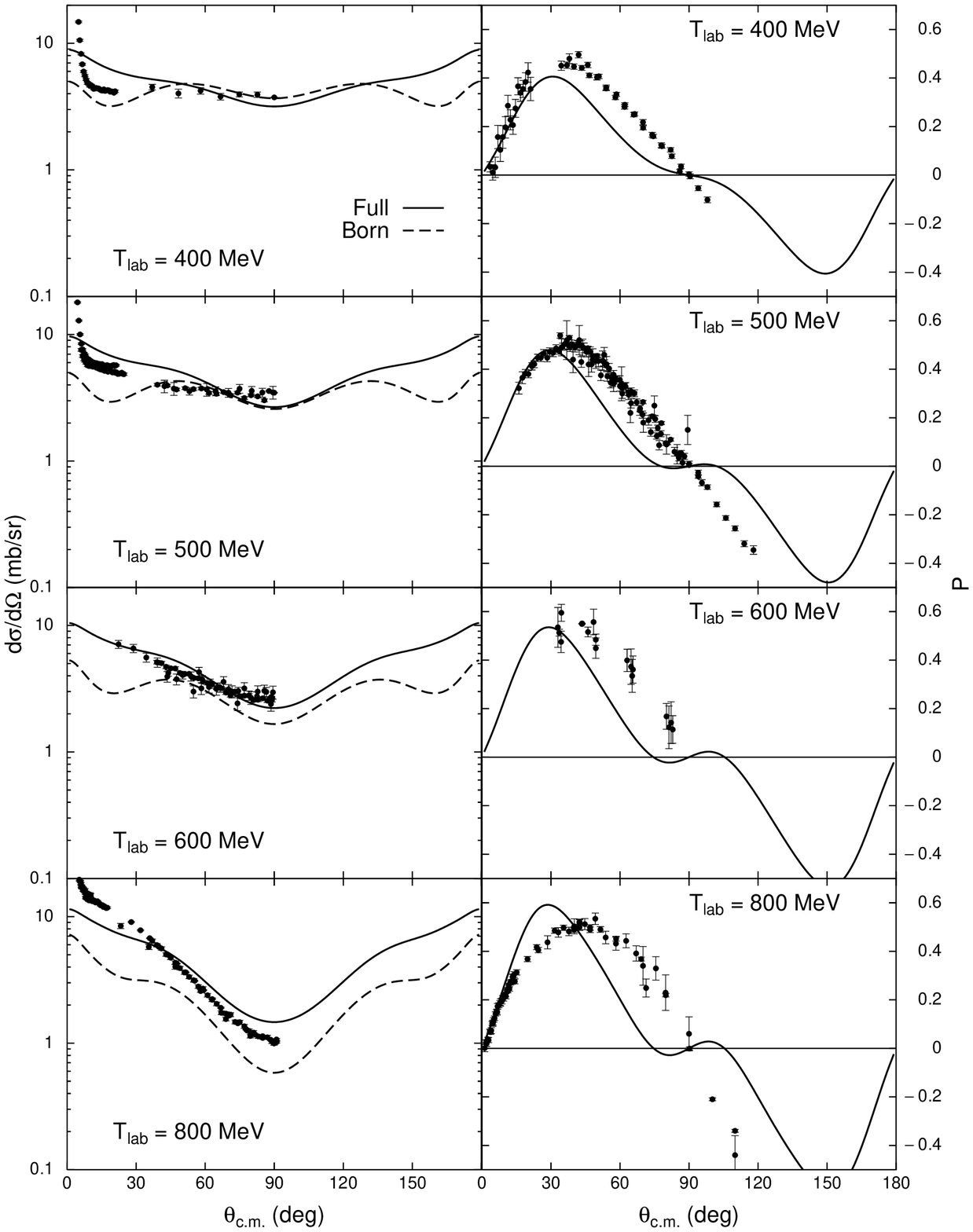}}
\vspace{0mm}
\caption{The same as Fig.\,2, but for the elastic $pp$ scattering
in model RGM-H. The Coulomb force is neglected.}
\label{fig3}
\end{minipage}
\end{center}
\end{figure}

Figures 2 and 3 compare with the experimental 
data \cite{SAID} the model predictions
of the elastic differential cross sections and the polarizations
for the the $np$ and $pp$ scattering
in the $T_{lab}=400\sim 800$ MeV range.
These observables in the lower energies are given
in Ref.\,\citen{FJ98}.
The model in Fig.\,2 is FSS for the $np$ scattering,
and that in Fig.\,3 is RGM-H for the $pp$ scattering.
The Coulomb force for the $pp$ scattering is neglected,
since it only affects the extreme
forward and backward angles in this high energy region. 
The solid curve indicates the predictions
obtained by solving the LS-RGM equation, while 
the dashed curve the results of the
Born approximation. The latter approximation 
is apparently inappropriate even at these high energies.
However, it is outstanding that the calculated Born invariant
amplitudes, leading to these cross sections in the Born
approximation, have almost the same order
of magnitude as the empirical amplitudes
determined from the phase-shift analysis \cite{SAID}.
Note that the polarization vanishes
in the Born approximation.
In these calculations no imaginary potential is introduced.
The theory overestimates the $np$ differential cross sections 
at backward angles. The $np$ polarization has an unpleasant
oscillation around $\theta_{c.m.}\sim 110^\circ$.  
There appears a symmetry in the $pp$ scattering because of the identity 
of two protons: The differential cross section becomes
symmetric with respect to $\theta_{c.m}=90^\circ$,
and the polarization is symmetric with an opposite sign.
The $pp$ polarization for $T_{lab}\geq 400$ MeV
shows an oscillation around $90^\circ$, which is not 
present in the experiment.
Except for these disagreements, the
characteristic behavior of the energy dependence
and the angular distribution is reasonably well reproduced
within the wide energy range up to 800 MeV.
These results indicate that the LS-RGM technique is
very useful for investigating the baryon-baryon
interaction above 300 MeV.

Figure 4 shows the five invariant amplitudes, $g_0(\theta) \sim
h_P(\theta)$, as a function of the c.m. angle $\theta$,
predicted by FSS for the $np$ scattering
at $T_{lab}=800$ MeV. The left column displays the real part,
while the right column the imaginary part.
The predictions by the Paris potential \cite{PARI} are
also shown in the dashed curve for comparison. 
In these calculations the partial waves
up to $J=8$ are included.\footnote{Actually $J_{max}=8$ is
not big enough for $T_{lab}=800~\hbox{MeV}$, as seen from
the small ripples of the solid and dashed curves
in Fig.\,4. The partial-wave contributions for $J > J_{max}$ 
from the Born amplitudes are added to obtain the results
in Figs.\,2 and 3.}
The dotted curve (Arndt) is the empirical value,
which is calculated from the solution
of the phase-shift analysis \cite{SP82} by using
the real part of the phase-shift parameters up to $J \leq 6$
and the partial-wave expansion of the invariant
amplitudes \eq{d2}.
It is clear that the most prominent disagreement between
the FSS prediction and the other predictions appears 
in the real part of the spin-independent central invariant 

%
%\begin{wrapfigure}{c}{\textwidth}
\begin{figure}[b]
\epsfxsize=\textwidth
\centerline{\epsfbox{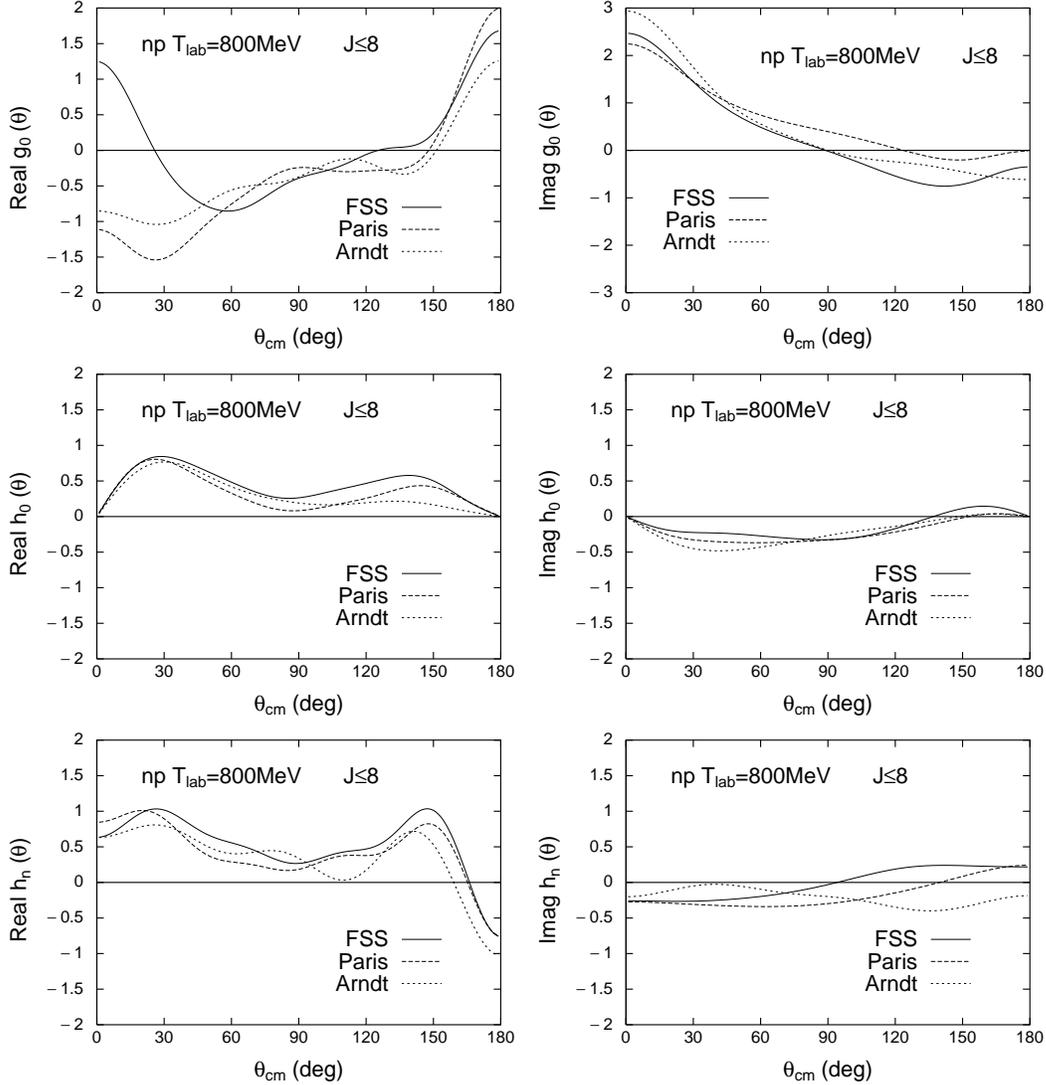}}
\vspace{0mm}
\caption{The real (left) and imaginary (right) parts
of the five invariant amplitudes, $g_0(\theta)
\sim h_P(\theta)$, for the $np$ elastic scattering
at $T_{lab}=800$ MeV, predicted by model FSS.
The dashed curve stands for predictions
by the Paris potential \protect\cite{PARI},
and the dotted curve those by the phase-shift analysis
SP82 by Arndt {\em et al.} \protect\cite{SP82}. 
The partial waves included are $J \leq 6$ for SP82,
and $J \leq 8$ otherwise.}
\label{fig4a}
%\end{wrapfigure}
\end{figure}
%
%\newpage
%
\addtocounter{figure}{-1}
%\begin{wrapfigure}{c}{\textwidth}
\begin{figure}[t]
\epsfxsize=\textwidth
\centerline{\epsfbox{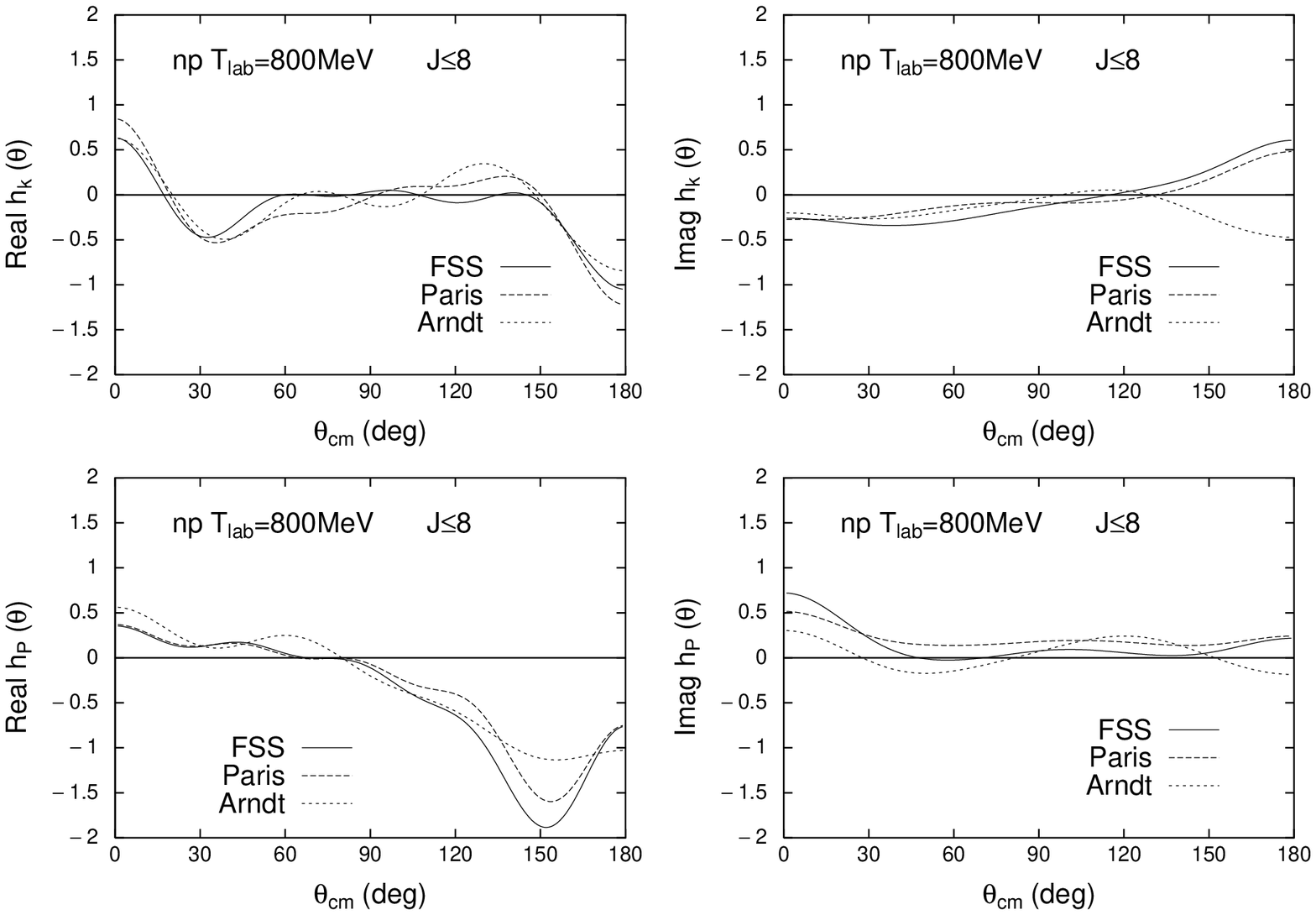}}
\vspace{0mm}
\caption{-continued}
\label{fig4b}
%\end{wrapfigure}
\end{figure}

\noindent
amplitude $\Re e~g_0(\theta)$ at the forward
angle. FSS predicts $\Re e~g(0) \sim 1.25$,
while the Paris potential and the phase-shift analysis
predict $\Re e~g(0) \sim -1$. 
If we use the $t^{eff} \rho$ prescription
discussed in $\S$\,2.4, these values correspond to the
nucleon s.p. potential in normal nuclear matter, $-17$ MeV for FSS
and $+14$ MeV for the latter two, as a contribution from the unlike
nucleons.\footnote{These are consistent with the numbers given
in Table VI. Namely, if we add up the $I=0$ contribution and
one-third of the $I=1$ contribution to the real part of
the s.p. potentials at 800 MeV, the potential depth
becomes $-17.4$ MeV for FSS, $15.5$ MeV for the Paris potential,
and $14.5$ MeV for SP99.}
We here find that the attractive behavior
of the FSS s.p. potentials around this energy region 
is related to the wrong sign of the real part
of the spin-independent central invariant
amplitude $\Re e~g_0(\theta)$ at the forward direction.
This difference does not impair the differential cross sections
very much, since the invariant amplitude has an appreciable magnitude
for the imaginary part as seen in Fig.\,4. This should, however,
affect some particular polarization observables,
and cause a large disagreement compared to experiment.

%\newpage

\section{Nucleon and hyperon single-particle
potentials in nuclear matter}

\begin{figure}[t]
\epsfxsize=\textwidth
\centerline{\epsfbox{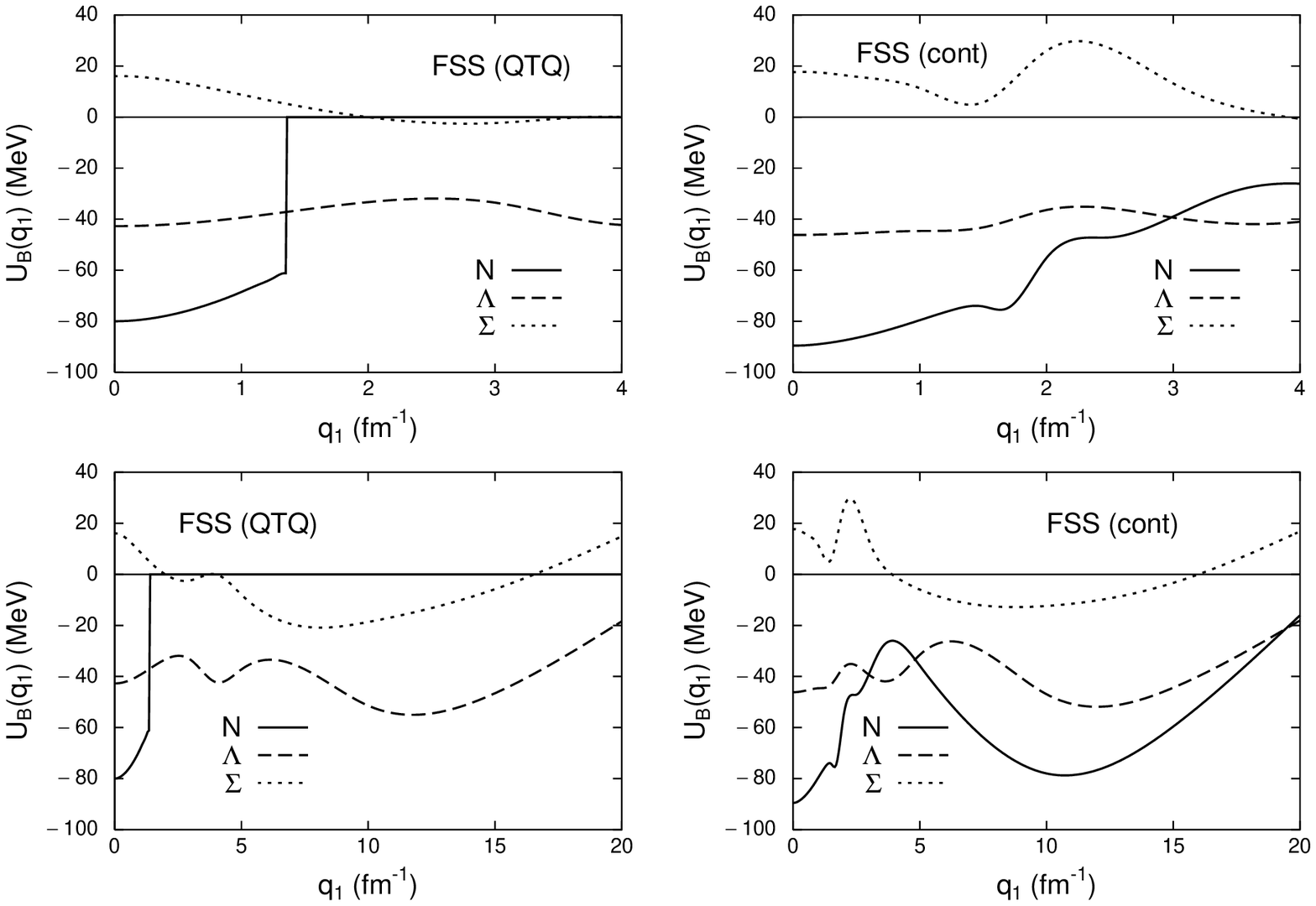}}
\vspace{0mm}
\caption{The nucleon and hyperon ($\Lambda$, $\Sigma$) s.p. potentials
predicted by the $G$-matrix calculation of model FSS.
The results in the left panels are obtained by using $QTQ$
prescription, while those in the right panels
by the continuous prescription for intermediate energy spectra.
The momentum interval
is $0 < q_1 < 4~\hbox{fm}^{-1}$ in the upper panels, and
$0 < q_1 < 20~\hbox{fm}^{-1}$ in the lower panels.
The Fermi momentum $k_F=1.35~\hbox{fm}^{-1}$ is assumed.}
\label{fig5}
\end{figure}

Figure 5 illustrates s.p. potentials $U_B(q_1)$, as a function 
of the incident momentum $q_1$, which are obtained for $B=N$,
$\Lambda$ and $\Sigma$ from
the $G$-matrix calculations with the model FSS.
The normal density with $k_F=1.35~\hbox{fm}^{-1}$ is assumed
for nuclear matter.
The left panels are the results with $QTQ$ prescription
for intermediate energy spectra, while the right panels are 
those with the continuous prescription. \cite{KO99}
The upper panels are for the momentum $q_1 < 4~\hbox{fm}^{-1}$,
while the lower panels for $q_1 < 20~\hbox{fm}^{-1}$.
In the $QTQ$ prescription, the self-consistency of the s.p. potentials
is not respected in the momentum region $q_1 >
k_F=1.35~\hbox{fm}^{-1}$. On the other hand, it is fully accounted
for in the continuous prescription.\footnote{In Ref.\,\citen{KO99}
we have assumed $U_B (q_1)=U_B (q_1=3.8~\hbox{fm}^{-1})$ for $q_1
\geq 3.8~\hbox{fm}^{-1}$, in order to avoid the unrealistic behavior
of the s.p. potentials in the high momentum region.
Here $q_1=3.8~\hbox{fm}^{-1}$ corresponds
to $T_{lab}=300~\hbox{MeV}$ in the $NN$ scattering.}
This implies that we first calculate $U_N(q_1)$,
and then $U_\Lambda(q_1)$ and $U_\Sigma(q_1)$ are determined
self-consistently by using the result of $U_N(q_1)$.
The partial waves included are for $J \leq 9$.
We can see from Fig.\,5 that the s.p. potentials predicted by
FSS is fairly strongly attractive in the momentum
interval $q_1=5 \sim 20~\hbox{fm}^{-1}$ for all the baryons.
In particular, $U_N(q_1)$ in the continuous prescription
becomes almost $-80$ MeV at $q_1=10~\hbox{fm}^{-1}$.  
This momentum interval corresponds to the incident
energy $T_{lab}=500~\hbox{MeV} \sim 8~\hbox{GeV}$ in
the $NN$ scattering (see Table III).
We therefore need to examine the invariant amplitudes carefully
in this energy region. 

Let us first examine whether the $t^{eff} \rho$~~prescription
discussed in $\S$\,2.4 is a good approximation to the
s.p. potentials predicted by the $G$-matrix calculation.
Table III shows such a comparison with respect to 
the nucleon s.p. potential $U_N(q_1)$ (in MeV) predicted
by the model FSS.
Here we find that the maximum value of the total angular
momentum $J_{max}=7$ is actually too small,
when the incident momentum

%\begin{table}[t]
\begin{wraptable}{c}{\textwidth}
\caption{Comparison of s.p. potential $U_N(q_1)$ (in MeV), obtained
by $t^{eff} \rho$~~prescription ($T$-matrix) and
by $G$-matrix calculation in the continuous choice \protect\cite{KO99}.
The model is FSS \protect\cite{FU96a,FU96b}.}
\label{table3}
\bigskip
(Real part)
%
%\bigskip
\begin{center}
\renewcommand{\arraystretch}{1.2}
\setlength{\tabcolsep}{3mm}
\begin{tabular}{rrrrr}
\hline
\hline
$T_{lab}$ & $q_1$ & $T$-matrix &
\multicolumn{2}{c}{$G$-matrix (cont.)} \\
\cline{3-5}
(MeV) & (fm$^{-1}$)
&  $J \leq 9$ &  $J \leq 9$ & $J \leq 7$ \\
\hline
187 & 3 & $-37.49$ & $-38.98$ & $-36.76$ \\
332 & 4 & $-27.81$ & $-26.08$ & $-23.75$ \\
518 & 5 & $-30.19$ & $-35.79$ & $-33.01$ \\
2074 & 10 & $-79.90$ & $-78.00$ & $-54.20$ \\
8295 & 20 & $-14.81$ & $-16.07$ & $-7.82$ \\
\hline
\end{tabular}
\end{center}
\bigskip
(Imaginary part)
%
%\bigskip
\begin{center}
\renewcommand{\arraystretch}{1.2}
\setlength{\tabcolsep}{3mm}
\begin{tabular}{rrrrr}
\hline
\hline
$T_{lab}$ & $q_1$ & $T$-matrix &
\multicolumn{2}{c}{$G$-matrix (cont.)} \\
\cline{3-5}
(MeV) & (fm$^{-1}$)
&  $J \leq 9$ &  $J \leq 9$ & $J \leq 7$ \\
\hline
187 & 3 & $-44.91$ & $-29.71$ & $-29.53$ \\
332 & 4 & $-52.88$ & $-39.64$ & $-39.43$ \\
518 & 5 & $-60.51$ & $-44.68$ & $-44.88$ \\
2074 & 10 & $-34.91$ & $-36.81$ & $-29.77$ \\
8295 & 20 & $-22.96$ & $-21.07$ & $-2.92$ \\
\hline
\end{tabular}
\end{center}
%\end{table}
\end{wraptable}

\begin{wraptable}{c}{\textwidth}
%\begin{table}[b]
\caption{Decomposition of $U_N(q_1)$ (in MeV) in
Table III to the isospin $I=0$ and $I=1$ components.
The total angular momentum included is $J \leq 9$.
The model is FSS \protect\cite{FU96a,FU96b}.}
\label{table4}
\bigskip
(Real part)
\begin{center}
\renewcommand{\arraystretch}{1.2}
\setlength{\tabcolsep}{3mm}
\begin{tabular}{rrrrrr}
\hline
\hline
$T_{lab}$ & $q_1$ &
\multicolumn{2}{c}{$I=0$} & \multicolumn{2}{c}{$I=1$} \\
\cline{3-6}
(MeV) & (fm$^{-1}$) & $T$-matrix &
$G$-matrix & $T$-matrix & $G$-matrix \\
\hline
187 & 3 & $-11.91$ & $-15.25$ & $-25.58$ & $-23.73$ \\
332 & 4 & $-5.98$ & $-7.53$ & $-21.83$ & $-18.55$ \\
518 & 5 & $-5.04$ & $-9.65$ & $-25.16$ & $-26.14$ \\
2074 & 10 & $-28.72$ & $-27.93$ & $-51.18$ & $-50.07$ \\
8295 & 20 & $-2.37$ & $-3.60$ & $-12.44$ & $-12.47$  \\
\hline
\end{tabular}
\end{center}
\bigskip
(Imaginary part)
\begin{center}
\renewcommand{\arraystretch}{1.2}
\setlength{\tabcolsep}{3mm}
\begin{tabular}{rrrrrr}
\hline
\hline
$T_{lab}$ & $q_1$ &
\multicolumn{2}{c}{$I=0$} & \multicolumn{2}{c}{$I=1$} \\
\cline{3-6}
(MeV) & (fm$^{-1}$) & $T$-matrix &
$G$-matrix & $T$-matrix & $G$-matrix \\
\hline
187 & 3 & $-22.02$ & $-15.85$ & $-22.89$ & $-13.86$ \\
332 & 4 & $-22.66$ & $-17.69$ & $-30.22$ & $-21.57$ \\
518 & 5 & $-22.21$ & $-19.69$ & $-38.30$ & $-24.99$ \\
2074 & 10 & $-15.27$ & $-14.83$ & $-19.63$ & $-21.99$ \\
8295 & 20 & $-20.62$ & $-18.78$ & $-2.34$ & $-2.29$  \\
\hline
\end{tabular}
\end{center}
%\end{table}
\end{wraptable}

\newpage

\noindent
$q_1 \geq 10~\hbox{fm}^{-1}$.
However, if we take the same $J_{max}$ in the two calculations,
the accuracy of the $t^{eff} \rho$~~prescription seems to be
fairly good even for the large momentum
around $q_1 \sim 10~\hbox{fm}^{-1}$.
Quite surprisingly, this approximation is very good even
at such a small energy as $T_{lab}=200$ MeV, as long as
the real part of the s.p. potential is concerned.
This agreement between the two prescriptions becomes clearer,
if we examine each isospin component with $I=0$ and $I=1$,
separately, as seen from Table IV.
As to the imaginary part of the s.p. potential,
the $t^{eff} \rho$~~prescription seems to overestimate
the values by the $G$-matrix calculation.

\begin{table}[t]
\caption{Reduction factors of the $S$-meson
central attraction due to the momentum-dependent $\bq^2$ term.}
\label{table5}
\bigskip
\begin{center}
\renewcommand{\arraystretch}{1.2}
\setlength{\tabcolsep}{6mm}
\begin{tabular}{crc}
\hline
\hline
$q \sim k_{cm}$ & $T_{lab}$ & $0.0243~q^2$ \\
\hline
2 & 332 MeV & 0.097 \\
3.1 & 800 MeV & 0.23 \\
5   & 2 GeV   & 0.61 \\
7.5 & 4.7 GeV & 1.37 \\
\hline
\end{tabular}
\end{center}
\end{table}

From this comparison and the behavior of the invariant amplitudes
in the preceding section, we have found that the attractive behavior
of $U_N(q_1)$ in the momentum interval $q_1=5 \sim
20~\hbox{fm}^{-1}$ in FSS is related to the wrong sign
(of the real part) of the spin-independent central invariant amplitude
at the forward direction. We expect that this situation is
common even for the $\Lambda$ and $\Sigma$ hyperons, and
it is a flaw of our present quark model (not only FSS,
but also RGM-H and RGM-F).
It should be mentioned that the quantitative aspect
of the present non-relativistic single-channel calculation in the
above momentum region is not entirely trustable.
The corresponding energy region of the $NN$ scattering
is already the relativistic energy region, where 
many inelastic channels are open.
Nevertheless, a repulsive behavior of the s.p. potential
at $q_1=5 \sim 20~\hbox{fm}^{-1}$ seems to be a mandatory
requirement even in the single-channel calculation,
since channel-coupling effects are expected
to work attractive to the s.p. potential.
Although this energy region may already be out of the applicability
of our non-relativistic quark model,
we need s.p. potentials for the intermediate states
with quite high momenta when the $G$-matrix calculation
is carried out in the continuous prescription. 

In order to solve this problem, we use an advantage
of our quark model that the effect of the short-range correlation
is rather moderate compared with that of the standard meson-exchange
potentials like the Paris potential.
Namely, an improvement of the Born amplitudes is clearly reflected
to the improvement of the solution of the LS-RGM equation.
In particular, the intermediate-range attraction from
the scalar-meson exchange has the Born kernel
\begin{eqnarray}
& & V^C(\bk, \bq)= - {g^2 \over \bk^2+m^2}
~\left[~1-{\bq^2 \over 2M^2}+{\bk^2 \over 8M^2}~\right]
\label{sp11}
\end{eqnarray}
with $\bk=\bq_f-\bq_i$ and $\bq=(\bq_f+\bq_i$)/2,
in the approximation up to the order of $(v/c)^2$.
Here $m$ and $M$ are the meson mass and the baryon mass, respectively.
So far we have used only the leading term 1 in the square bracket
of \eq{sp11} and neglected $\bq^2$ and $\bk^2$ terms.
A dominant contribution to the spin-independent central invariant
amplitude at the forward angle comes from the $\bq^2$ term,
which becomes important in the high-energy region. 
In fact, if we set $\bk^2 \rightarrow -m^2$ as usual,
modify the square bracket of \eq{sp11} as
\begin{eqnarray}
& & \left[~1-{\bq^2 \over 2M^2}+{\bk^2 \over 8M^2}~\right]
\longrightarrow \left[~1-{\bq^2 \over 2M^2}
-{m^2 \over 8M^2}~\right] \nonumber \\
& & = \left(1-{m^2 \over 8M^2}\right)
\left[~1-{ 1 \over 1-m^2/8M^2}{\bq^2 \over 2M^2}
~\right]\ \ ,
\label{sp12}
\end{eqnarray}
and redefine $g^2$ including the $\left(1-m^2/8M^2 \right)$ factor,
we find the momentum dependence like $1-0.0243~\bq^2$ for the
mass $mc^2=800$ MeV of the $\epsilon$ meson in FSS.
Here $|\bq|$ is in units of $\hbox{fm}^{-1}$.
This non-static term of the $S$ mesons plays a role to reduce
the strength of the intermediate-range attraction
by about 20 $\%$ at 800 MeV. (See Table V.)
Actually the momentum $\bq$ is not directly related to
the total energy, nor the direct Born term is good enough
to discuss the reduction of the central attraction at
higher energies. We also have contributions
from the inherent zero-point oscillation
of the cluster wave functions and those from the
quark-exchange kernel, since our EMEP are acting between quarks.
Nevertheless, the discussion here is still valid, since the dominant
contribution to the intermediate attraction is the direct
term of the $\epsilon$-meson exchange potential.
Bryan-Scott \cite{BR67} carefully examined
these $\bq^2$ momentum-dependent terms in the S-meson
and V-meson exchange potentials.
They found that the inclusion
of these terms has a favorable effect of making
the non-relativistic approximation uniform to 
order $\bq^2$, although the main effect is almost compensated
for by a slight change in the coupling constants.
Since these terms are included in the Paris potential \cite{PARI}
and all the Nijmegen soft-core potentials \cite{NSC89,NSC97},
it would be useful to incorporate these terms in our quark model,
in order to describe correctly the asymptotic behavior
of the s.p. potentials in the high-momentum region.

%\begin{wraptable}{c}{\textwidth}
\begin{table}[t]
\caption{Comparison of $U_N(q_1)$ (in MeV) in
the $t^{eff} \rho$ prescription and its decomposition
to the isospin $I=0$ and $I=1$ components,
with other models (Paris \protect\cite{PARI}) and
experiment (SP99 \protect\cite{SAID}).
The result by a new model with
the momentum-dependent $\bq^2$ term (present) is also shown.}
\label{table6}
\bigskip
\begin{center}
\renewcommand{\arraystretch}{1.2}
\setlength{\tabcolsep}{3mm}
\vspace{0mm}
\begin{tabular}{crrrrrr}
\hline
\hline
$T_{lab}$ (MeV) & \multicolumn{6}{c}{FSS ($J \leq 8$)} \\
\cline{2-7}
or $q_1$ (fm$^{-1}$) & \multicolumn{2}{c}{$I=0$} &
\multicolumn{2}{c}{$I=1$} &\multicolumn{2}{c}{total} \\
\cline{2-7}
 & real & imag & real & imag & real & imag \\
\hline
200 & $-11.43$ & $-22.08$ & $-25.47$ & $-23.48$ & $-36.90$ & $-45.56$
\\
400 & $-5.11$ & $-22.59$ & $-23.81$ & $-33.51$ & $-28.92$ & $-56.11$
\\
800 & $-5.58$ & $-19.64$ & $-35.41$ & $-44.37$ & $-40.99$ & $-64.01$
\\
$q_1$=10 & $-17.36$ & $-10.48$ & $-53.92$ & $-19.57$ & $-71.27$ &
$-30.05$ \\
$q_1$=20 & $-7.04$ & $-1.83$ & $-12.52$ & $-2.34$ & $-19.56$ & $-4.16$ \\
\hline
\hline
$T_{lab}$ (MeV) & \multicolumn{6}{c}{present ($J \leq 8$)} \\
\cline{2-7}
or $q_1$ (fm$^{-1}$) & \multicolumn{2}{c}{$I=0$} &
\multicolumn{2}{c}{$I=1$} &\multicolumn{2}{c}{total} \\
\cline{2-7}
 & real & imag & real & imag & real & imag \\
\hline
200 & $-9.40$ & $-16.72$ & $-26.02$ & $-20.72$ & $-35.42$ & $-37.45$
\\
400 & $-1.90$ & $-16.11$ & $-13.41$ & $-26.58$ & $-15.31$ & $-42.70$
\\
800 & $4.18$ & $-15.48$ & $-4.24$ & $-38.18$ & $-0.06$ & $-53.63$
\\
$q_1$=10 & $13.44$ & $-11.52$ & $29.23$ & $-50.58$ & $42.67$ &
$-62.10$ \\
$q_1$=20 & $39.4$ & $-47.3$ & $176$ & $-150$ & $215$ & $-197$ \\
\hline
\hline
$T_{lab}$ (MeV) & \multicolumn{6}{c}{Paris potential ($J \leq 8$)} \\
\cline{2-7}
or $q_1$ (fm$^{-1}$) & \multicolumn{2}{c}{$I=0$} &
\multicolumn{2}{c}{$I=1$} &\multicolumn{2}{c}{total} \\
\cline{2-7}
 & real & imag & real & imag & real & imag \\
\hline
200 & $-9.23$ & $-17.74$ & $-27.37$ & $-19.50$ & $-36.60$ & $-37.25$
\\
400 & $0.65$ & $-16.87$ & $-10.76$ & $-25.09$ & $-10.11$ & $-41.96$
\\
800 & $10.28$ & $-18.97$ & $15.68$ & $-36.98$ & $25.96$ & $-55.95$
\\
$q_1$=10 & $25.67$ & $-34.83$ & $78.49$ & $-89.93$ & $104$ &
$-125$ \\
$q_1$=20 & $32.78$ & $-83.78$ & $143$ & $-266$ & $176$ & $-350$ \\
\hline
\hline
$T_{lab}$ (MeV) & \multicolumn{6}{c}{SP99 (Phase shift analysis)
($J \leq 7$)} \\
\cline{2-7}
 & \multicolumn{2}{c}{$I=0$} &
\multicolumn{2}{c}{$I=1$} &\multicolumn{2}{c}{total} \\
\cline{2-7}
 & real & imag & real & imag & real & imag \\
\hline
200 & $-8.56$ & $-16.37$ & $-25.68$ & $-19.76$ & $-34.24$ & $-36.13$
\\
400 & $3.12$ & $-15.08$ & $-15.87$ & $-30.13$ & $-12.76$ & $-45.21$
\\
800 & $13.28$ & $-16.51$ & $3.53$ & $-74.20$ & $16.82$ & $-90.71$
\\
\hline
\end{tabular}
\end{center}
%\end{wraptable}
\end{table}

As an example of the quark model
with the momentum-dependent $\bq^2$ term,
we show in Table VI the result
of the s.p. potential $U_N(q_1)$ in the $t^{eff} \rho$ prescription,
calculated by a new model (present).
In this new model V-mesons are also incorporated as the
EMEP acting between quarks. The phase-shift parameters
of the $np$ scattering is largely improved in comparison
with those of FSS.
The details of this new model will be published elsewhere.
In Table VI predictions of the Paris potential \cite{PARI}
and of the phase-shift analysis SP99 \cite{SAID},
together with the decomposition
to $I=0$ and $I=1$ contributions, are also shown
for comparison.
We find that the new model is still slightly too attractive
around $T_{lab} \sim 800~\hbox{MeV}$, but the flaw of FSS
with too attractive asymptotic behavior
in the high-momentum region is clearly removed.

%\newpage

\section{Summary}

In the quark-model study of the nucleon-nucleon ($NN$) and
hyperon-nucleon ($YN$) interactions,
a variational method is usually used to solve
an integro-differ-
\break
ential equation
formulated in the $(3q)$-$(3q)$ resonating-group method (RGM).
In this paper we have developed a Lippmann-Schwinger formalism
for the $(3q)$-$(3q)$ RGM as an alternative method
to the improved variational method \cite{KA77}.
The basic equation, which we call the LS-RGM equation, is
completely equivalent to the standard RGM equation in the
coordinate representation on the energy-shell.
An advantage of solving the RGM equation
in the momentum representation is to avoid rapid oscillation
of the relative wave functions at higher energies,
which leads to the strong cancellation of
Gaussian trial functions in the improved variational method.
This feature of the LS-RGM formalism naturally makes it
possible to obtain an accurate $S$-matrix even
for the relativistic energies $T_{lab} \geq 350$
MeV, where many inelastic channels open.
Since the $S$-matrix is very accurate also in the low-energy region,
an extension to the coupled-channel systems
with different threshold energies is very successful. 

In this formulation the Born kernel is analytically calculated
for all pieces of direct and exchange terms,
which are composed of the kinetic-energy term
and various pieces of quark-quark ($qq$) interactions. 
The $qq$ interactions are further divided
into the phenomenological confinement potential, 
a color analogue of the Fermi-Breit (FB) interaction,
and the effective meson-exchange
potentials (EMEP) acting between quarks.
The Born kernel is then decomposed into partial-wave
components and the resultant LS-RGM equation
is solved by using the techniques developed by
Noyes \cite{NO65} and Kowalski \cite{KO65}.
Since calculations are always carried out in the momentum
representation, the present formalism has no difficulty
to incorporate the momentum-dependent $qq$ interaction such as
the momentum-dependent Breit retardation term
of the FB interaction, the higher-order terms of the central
scalar-meson and vector-meson exchange potentials,
and the quadratic $LS$ force.   
A convenient transformation formula to derive spatial functions
for the direct and exchange kernels is given
for a very general type of two-body interactions.
The numerical evaluation of the partial-wave Born kernel
is also best suited to convert the LS-RGM equation
to the $G$-matrix (Bethe-Goldstone) equation \cite{KO99},
in which the Pauli principle is treated exactly at the baryon level.

The accuracy of the LS-RGM formalism has been examined
in the $\Lambda N$-$\Sigma N (I=1/2)$ coupled-channel system.
In this system the $\hbox{}^3S_1$-$\hbox{}^3D_1$ coupling
caused by the very strong one-pion tensor force yields a prominent
cusp structure for the $\Lambda N$ phase-shift parameters
at the $\Sigma N$ threshold. On the other hand, a resonance
appears either in $\Lambda N$ $\hbox{}^1P_1$ state
or in $\Sigma N$ $\hbox{}^3P_1$ state by the strong
effect of the antisymmetric $LS$ force ($LS^{(-)}$ force) originating
from the FB interaction. The behavior of these resonances
is rather sensitive to the characteristics of the model,
particularly to the strength of the central attraction
of the $\Sigma N (I=1/2)$ channel.
We have examined the model RGM-F and obtained a satisfactory
agreement of the phase-shift parameters between the LS-RGM method
and the improved variational method,
in the wide momentum region $p_{lab}=0 \sim 1~\hbox{GeV}/c$.

Using the same parameter set determined
at $T_{lab}=0 \sim 250~\hbox{MeV}$, we have
extended our calculation of the differential cross sections and the
polarization for the $NN$ scattering to the intermediate
energies $T_{lab}=400 \sim 800~\hbox{MeV}$.
Although agreement with the experimental data
becomes gradually worse as the energy becomes higher,
the characteristic behavior of the energy dependence
and the angular distribution of these observables
for the elastic $np$ and $pp$ scattering are reasonably reproduced
within this energy range.
In particular, the invariant amplitudes predicted by model FSS
for the $np$ scattering at $T_{lab}=800$ MeV reproduce reasonably well 
the empirical invariant amplitudes determined
from the phase shift analysis, except for a few typical disagreements.
The most prominent disagreement appears in the real part
of the spin-independent central invariant
amplitude $g_0(\theta)$ at the forward angles $\theta \sim 0$.
This amplitude is related to the single-particle (s.p.) potential
obtained from the $G$-matrix calculation
through the $t^{eff}\rho$ prescription,
when the incident momentum is high.
The wrong sign of $\Re e g_0(\theta)$ at $\theta=0$ in
our model is correlated to too attractive s.p. potentials
in the momentum region $q_1=5 \sim 20~\hbox{fm}^{-1}$.

We have also examined the accuracy of the $t^{eff}\rho$ prescription
by using the $G$-matrix solution \cite{KO99} and
the present LS-RGM formalism.
This prescription is a good approximation
for the real part of the s.p. potentials in the energy
region from $T_{lab}=200$ MeV to several GeV ($q_1=3~\hbox{fm}^{-1}
\sim 20~\hbox{fm}^{-1}$), as long as the maximum value of the
angular-momentum cut for partial waves is commonly taken.
On the other hand, the imaginary parts of the s.p. potentials
are usually overestimated in the $t^{eff} \rho$~~prescription.

Perhaps the most striking feature of the quark model found in this
investigation is a moderate effect of the short-range correlation.
In the standard meson-exchange models, the observed phase-shifts
are reproduced as a cancellation of very strong repulsive
and attractive local potentials, and the Born amplitudes of the
Paris potential, for example, are one or two order
of magnitude large, compared to the empirical invariant amplitudes.
On the other hand, the short-range repulsion in the quark model
originates mainly from the nonlocal kernel of the color-magnetic 
term of the FB interaction. 
Born amplitudes of the quark model therefore have almost the same order
of magnitude as the empirical amplitudes obtained by solving
the LS-RGM equation. This implies that the short-range correlation
in the quark model is rather moderate compared with the
meson-exchange models.
It can also be seen in our recent quark-model study
of the s.p. spin-orbit potentials for the nucleon and hyperons.
In Ref.\,\citen{LS99} we have calculated the strength factor,
$S_B$, of the s.p. spin-orbit potentials by using the $G$-matrix
solutions and found that $S_N$ does not obtain much effect
of the short-range correlation, on the contrary to the standard
potential models like the Reid soft-core potential with
the strong short-range repulsive core.
Since the Born amplitudes in the quark model reflect
rather faithfully characteristics of the LS-RGM solution,
it is easy to find missing ingredients that impair the model.
In fact, we have discussed that the wrong sign of the invariant
amplitude $\Re e~g_0(\theta)$ in our quark model
is related to our neglect of higher-order
momentum-dependent central term of the scalar-meson
exchange EMEP. A preliminary result of a new model, which incorporates
this term as well as the vector-meson EMEP, shows that the
s.p. potentials in the $t^{eff}\rho$ prescription have
a correct repulsive behavior in the asymptotic momentum region. 
The details of this model will be given in a forthcoming paper.

%\newpage

\appendix

\section{A transformation formula to the Born kernel
for momentum-dependent two-body interactions}

As to the general procedure how to calculate the RGM kernel
for two-cluster systems composed of $s$-shell clusters,
Refs.\,\citen{SU83} and \citen{KI94} should be referred to.
Here we use the same notation as Ref.\,\citen{KI94} and
give a convenient formula to calculate the Born kernel
for two-body interactions with momentum dependence.

Suppose a two-body interaction of the RGM Hamiltonian is given
by $v_{ij}=u_{ij} w_{ij}$ with the spatial
part $u=u(\br,\partial /\partial \br)$
and the spin-flavor-color part $w_{ij}=w_{ij}^{SF} w_{ij}^C$.
The full Born kernel for this interaction is defined
by
\begin{equation}
\langle\,e^{i \bq_f \cdot \br}~\phi\,\vert
\sum_{i<j}^6 v_{ij} \CA^\prime \,\vert
\,e^{i \bq_{\,i}\cdot \br}~\phi \rangle\ \ ,
\label{a1}
\end{equation}
where $\phi=\phi^{space}\,\xi$ with $\xi=\xi^{SF} \xi^C$ is
the harmonic-oscillator (h.o.) internal
wave function of the $(3q)$-$(3q)$ system
and $\CA^\prime$ is the antisymmetrization
operator between two clusters.
This expression is reduced to the form \eq{fm15} by the use of the
double coset expansion $\CA^\prime \rightarrow (1/2)(1-9P_{36})
(1-P_0)$ with $P_0=P_{14}P_{25}P_{36}$,
yielding the basic Born kernel Eqs.\,(\ref{fm23}) and (\ref{fm27})
with
\begin{equation}
M(\bq_f, \bq_i)=\sum_{x \CT} X_{x \CT}
~M_{x \CT}(\bq_f, \bq_i)\ \ .
\label{a2}
\end{equation}
Here the superscript $\Omega$ for specifying the type
of the interaction is omitted for simplicity,
and the sum over $x$ is only for $x=0$ and 1. 
Furthermore, $X_{x \CT}$ is the spin-flavor-color factor
defined through
\begin{eqnarray}
X_{x \CT} & = & C_x \langle z_x~\xi\,|\,\sum_{i<j}^\CT w_{ij}
\,|\,\xi \rangle \nonumber \\
& = & \left\{
\begin{array}{c}
X^C_{0 \CT} \langle~\xi^{SF}\,|\,\sum_{i<j}^\CT w_{ij}^{SF}
\,|\,\xi^{SF}\,\rangle \\ [3mm]
(-9) X^C_{1 \CT} \langle~P_{36}~\xi^{SF}\,|
\,\sum_{i<j}^\CT w_{ij}^{SF}
\,|\,\xi^{SF}\,\rangle \\
\end{array} \right.
\quad \hbox{for} \quad
x=\left\{ \begin{array}{c}
0 \\ [3mm]
1 \\ \end{array} \right.,
\label{a3}
\end{eqnarray}
where $x=0$ with $z_0=1$ and $C_0=1$ corresponds to the direct term
and $x=1$ with $z_1=P_{36}$ and $C_1=-9$ the (one-quark) exchange term.
The suffix $\CT$ stands for the interaction type $\CT=E,
~S,~S^\prime,~D_+$ or $D_-$, for which the $(i,j)$ pairs are
properly selected.
The color-factors $X^C_{x \CT}=\langle z_x~\xi^C\,|
\,w^C_{ij}\,|\,\xi^C \rangle$ for each $(i,j) \in \CT$ are
given as follows. For the quark sector
with $w^C_{ij}=(1/4)(\lambda^C_i\lambda^C_j)$,
$X^C_{0E}=-(2/3)$,
$X^C_{0D_+}=0$, and $X^C_{1E}=X^C_{0S}=X^C_{0S^\prime}=-(2/9)$,
$X^C_{1D_+}=1/9$, $X^C_{1D_-}=4/9$.
For the EMEP sector with $w^C_{ij}=1$, $X^C_{0 \CT}=1$ and
$X^C_{1\CT}=1/3$.
The spin-flavor-color factor for the exchange normalization kernel
is defined by $X_N=(-3) \langle P_{36}~\xi^{SF}\,|
\,\xi^{SF}\,\rangle$.
We also need $X_K=24 \langle P_{36}~\xi^{SF}\,|\,Y(6)-Y(5)\,|
\,\xi^{SF}\,\rangle$ for the exchange kinetic-energy kernel
of the $YN$ systems.
On the other hand, the spatial part of the Born kernel is
defined by
\begin{equation}
M_{x \CT}(\bq_f, \bq_i)
=\langle\,z_x~e^{i \bq_f \cdot \br}\,\phi^{space}\,\vert
\,u_{ij}\,\vert
\,e^{i \bq_{\,i}\cdot \br}\,\phi^{space} \rangle
\quad \hbox{with} \quad (i,j) \in \CT\ \ ,
\label{a4}
\end{equation}
which we now evaluate.

The standard procedure is to use the h.o. generating function
\begin{equation}
A_\gamma(\br, \bz)=\left({2\gamma \over \pi}\right)^{{3 \over 4}}
~e^{-\gamma\left(\br-\bz/\sqrt{\gamma}\right)^2
+\bz^2/2}\ \ ,
\label{a5}
\end{equation}
and first to calculate the so-called complex GCM kernel
defined by
\begin{equation}
I_{x \CT}(\bz, \bz^\prime)
=\langle\,z_x~A_\gamma(\br, \bz)~\phi^{space}\,\vert
\,u_{ij}\,\vert
\,A_\gamma(\br, \bz^\prime)~\phi^{space} \rangle\ \ .
\label{a6}
\end{equation}
This GCM kernel is divided into the general form of the norm kernel
and the interaction function $\CT(\bz,\bz^\prime)$:
\begin{eqnarray}
I_{x \CT}(\bz, \bz^\prime)
& = & e^{\left(1-{x \over \mu}\right) \bz \cdot \bz^\prime}
~\CT(\bz, \bz^\prime)\ \ ,\nonumber \\
\CT(\bz,\bz^\prime) & = & \langle (0s)_{\bS_\alpha}|
(0s)_{\bS_\beta^\prime} \rangle^{-1}
\langle (0s)_{\bS_\gamma}|
(0s)_{\bS_\delta^\prime} \rangle^{-1} \nonumber \\
& & \times
\langle (0s)_{\bS_\alpha}(0s)_{\bS_\gamma}\,|\,u\,|
\,(0s)_{\bS_\beta^\prime}(0s)_{\bS_\delta^\prime}
\rangle\ \ .
\label{a7}
\end{eqnarray}
Here $(0s)_{\bS}$ stands for a localized $(0s)$ wave function
around $\bx=\bS$, and $\bS_\alpha$ etc. are the generator coordinates
specifying the position of clusters \cite{KI94}.
For a simple Gaussian two-body
interaction $u(\br) \sim e^{-\kappa \br^2}$,
$\CT(\bz,\bz^\prime)$ is given by
\begin{equation}
\CT(\bz, \bz^\prime)=\left({\nu \over \nu+\kappa}\right)^{{3 \over 2}}
~e^{-{\lambda \over 2}\left(p\bz^*+q\bz^\prime\right)^2}
\times (polynomial~terms)\ \ ,
\label{a8}
\end{equation}
where $\nu=1/2b^2$ is the h.o. constant of the $(0s)$ clusters
and $\lambda$ is given by $\lambda=(1/2\mu)(\kappa/(\nu+\kappa))$
with $\gamma=\mu \nu$ and $\mu=3\cdot 3/(3+3)=3/2$.
The parameters $p$ and $q$ are 0, 1 or $-1$, depending on $\CT$,
which are explicitly given in TABLE II of Ref.\,\citen{KI94}.
The transformation to the Born kernel is achieved through
\begin{eqnarray}
M_{x \CT}(\bq_f,\bq_i) & = & \left({\gamma \over 2\pi}\right)
^{3 \over 2}e^{{1 \over 4\gamma}(\bq_f^2+\bq_i^2)}
\int d \ba~d \bb~e^{-{\gamma \over 2}(\ba^2+\bb^2)}
~e^{-i\bq_f \cdot \ba+i\bq_i \cdot \bb} \nonumber \\
& & \times I_{x \CT}(\sqrt{\gamma}\ba, \sqrt{\gamma}\bb)\ \ .
\label{a9}
\end{eqnarray}

When $u$ involves momentum-dependence, it is convenient
to use the momentum representation and write
\begin{equation}
\langle\,\bp\,\vert u\,\vert\,\bp^\prime\,\rangle 
={1 \over (2\pi)^3}\,u(\bk^\prime, \bq^\prime)
\quad \hbox{with} \quad
\bk^\prime=\bp-\bp^\prime,~\bq^\prime={1 \over 2}
(\bp+\bp^\prime)\ \ .
\label{a10}
\end{equation}
Then the interaction function in \eq{a7} is given by
\begin{eqnarray}
\CT(\bz,\bz^\prime) & = & {1 \over (2\pi)^3}\,\left({1 \over \pi \nu}
\right)^{3 \over 2} \int d \bk^\prime~d \bq^\prime
~e^{-{1 \over \nu}\left({\bq^\prime}^2
+{1 \over 4}{\bk^\prime}^2 \right)}
~e^{{1 \over 4}\nu\br^2
+i\bq^\prime \cdot \br}~e^{i \bk^\prime \cdot \bX}
~u(\bk^\prime, \bq^\prime) \nonumber \\
& & \hbox{with} \quad
\br=-{1 \over \sqrt{\gamma}}(p\bz^*-q\bz^\prime),
~\bX= -{1 \over 2\sqrt{\gamma}}(p\bz^*+q\bz^\prime)\ \ .
\label{a11}
\end{eqnarray}
If we use \eq{a11} in \eq{a9}, the integration
over $\ba$ and $\bb$ can be carried out and we obtain the following
formula after some rearrangement of terms:
\begin{eqnarray}
M_{x \CT}(\bq_f,\bq_i) & = & M_x^N(\bq_f,\bq_i)
~\left({1 \over 2\pi}\right)^3 \int d \bk^\prime
\exp \left\{-\left(1+{\widetilde{\alpha} \over 2 \mu}\right)
{1 \over 4 \nu}
\bk^{\prime 2}-{1 \over 2 \sqrt{\gamma}} \bV \bk^\prime \right\}
\nonumber \\
& & \times \left({1 \over \pi \nu} {1 \over 1-\alpha/2\mu}
\right)^{3/2} \int d \bq^\prime
\exp \left\{-{1 \over \nu}{1 \over 1-\alpha/2\mu}
\bq^{\prime 2} \right\} \nonumber \\
& & \times u \left(\bk^\prime, \bq^\prime
-{\varepsilon \over 4 \mu}\bk^\prime
-{\nu \over 2\sqrt{\gamma}}\bA\right)\ \ .
\label{a12}
\end{eqnarray}
Here $M_x^N(\bq_f,\bq_i)$ is the normalization kernel given by
\begin{equation}
M_x^N(\bq_f,\bq_i)=\left({2\pi \over \gamma}
{1 \over 1-\tau^2}\right)^{3/2}
\exp \left\{-{ 1 \over 2\gamma}\left({1-\tau \over 1+\tau}\bq^2
+{1+\tau \over 1-\tau}{1 \over 4} \bk^2 \right)\right\}\ ,
\label{a13}
\end{equation}
with $\tau=1-x/\mu$, $\bk=\bq_f-\bq_i$ and $\bq=(\bq_f+\bq_i)/2$,
and the coefficients appearing in \eq{a12}
are defined by 
\begin{eqnarray}
\alpha & = & {p^2+q^2-2\tau pq \over 1-\tau^2}\ \ ,\quad
~\widetilde{\alpha}={p^2+q^2+2\tau pq \over 1-\tau^2}\ \ ,\quad
~\varepsilon={p^2-q^2 \over 1-\tau^2}\ \ , \nonumber \\
\bV & = & {1 \over \sqrt{\gamma}}\left({p-q \over 1+\tau}\bq
+{p+q \over 1-\tau}{1 \over 2} \bk\right),
~\bA={1 \over \sqrt{\gamma}}\left({p+q \over 1+\tau}\bq
+{p-q \over 1-\tau}{1 \over 2} \bk\right).
\label{a14}
\end{eqnarray}

Let us specialize the two-body interaction to the Yukawa function
and the Gaussian function:
\begin{eqnarray}
& & u(\bk, \bq)=\left\{
\begin{array}{c}
\displaystyle{{4\pi \over \bk^2+m^2}}~\widetilde{u}(\bk, \bq)
\\ [5mm]
\left({\pi \over \kappa}\right)^{3 \over 2}
e^{-{1 \over 4\kappa} \bk^2}~\widetilde{u}(\bk, \bq) \\
\end{array} \right.\ \ .
\label{a15}
\end{eqnarray}
Here $\widetilde{u}(\bk, \bq)$ stands for a polynomial function
of $\bk$ and $\bq$, and the degree of the polynomial is
usually at most the second order in $\bq$.
The Born kernel for the Yukawa function is calculated from
the formula for the Gaussian function through the integral
representation
\begin{equation}
{4\pi \over \bk^2+m^2}
=m {1 \over \sqrt{\pi}} \int_0^\infty
d u~e^{-{1 \over u^2}}
~\left({\pi \over \kappa}\right)^{3 \over 2}
e^{-{1 \over 4\kappa}\bk^2}
\quad \hbox{with} \quad \kappa=\left({mu \over 2}\right)^2\ .
\label{a16}
\end{equation}
The resultant Born kernel is expressed
by generalized Dawson's integrals $\widetilde{h}_n(x)$ and
modified Yukawa functions $\widetilde{{\cal Y}}_\alpha(x),
~\widetilde{{\cal Z}}_\alpha(x)$ etc., which are given
by the error function of the imaginary argument:
\begin{eqnarray}
& & \widetilde{h}_n(x)=(2n+1)\,e^{-x^2} \int_0^1 e^{x^2 t^2}\,t^{2n}
~d t=(2n+1)\,{e^{-x^2} \over x^{2n+1}} \int_0^x e^{u^2}\,u^{2n}
~d u\ \ , \nonumber \\
& & \widetilde{{\cal Y}}_\alpha(x)=e^{\alpha-x^2}
\int_0^1 e^{-{\alpha \over t^2}+x^2 t^2}~d t\ \ , \nonumber \\
& & \widetilde{{\cal Z}}_\alpha(x)=e^{\alpha-x^2}
\int_0^1 e^{-{\alpha \over t^2}+x^2 t^2}~t^4~d t\ \ .
\label{a17}
\end{eqnarray}
For the Gaussian kernel \eq{a12} is reduced into
\begin{equation}
M_{x \CT}(\bq_f,\bq_i)=M_x^N(\bq_f,\bq_i)
\left({\nu \over \nu+\kappa}
{1 \over 1+\lambda \widetilde{\alpha}}\right)^{3 \over 2}
\exp \left\{{1 \over 2}
{\lambda \over 1+\lambda \widetilde{\alpha}} \bV^2 \right\}
~\CP(\bq_f,\bq_i)\ \ ,
\label{a18}
\end{equation}
where the polynomial part is given by
\begin{eqnarray}
\CP(\bq_f,\bq_i) & = & \left({1+\lambda \widetilde{\alpha} \over
8\pi \gamma \lambda} \right)^{3 \over 2}
\int \bk^\prime~\exp \left\{-{1+\lambda \widetilde{\alpha} \over
8\gamma \lambda}~{\bk^\prime}^2 \right\} \nonumber \\
& & \times \left({1 \over \pi \nu} {1 \over 1-\alpha/2\mu}
\right)^{3/2} \int d \bq^\prime
\exp \left\{-{1 \over \nu}{1 \over 1-\alpha/2\mu}
\bq^{\prime 2} \right\} \nonumber \\
& & \times \widetilde{u} \left(\bk^\prime+2\sqrt{\gamma}
\widetilde{\bV}, \bq^\prime
-{\varepsilon \over 4 \mu}\bk^\prime
+{\nu \over 2\sqrt{\gamma}}\bW\right)\ \ .
\label{a19}
\end{eqnarray}
Here $\widetilde{u}(\bk^\prime, \bq^\prime)
=\widetilde{u}(-\bk^\prime, -\bq^\prime)$ and
the simplified notation
\begin{equation}
\widetilde{\bV}={\lambda \over 1+\lambda \widetilde{\alpha}}
~\bV\ \ ,\quad \bW=\bA-\varepsilon \widetilde{\bV}
\label{a20}
\end{equation}
are used.
If $u(\bk, \bq)$ does not involve $\bq$-dependence,
the $\bq^\prime$ integral is carried out in \eq{a19} and
we obtain
\begin{equation}
\CP(\bq_f,\bq_i)=\left({1+\lambda \widetilde{\alpha} \over
8\pi \gamma \lambda} \right)^{3 \over 2}
\int \bk^\prime~\exp \left\{-{1+\lambda \widetilde{\alpha} \over
8\gamma \lambda}~{\bk^\prime}^2 \right\}
~\widetilde{u} \left(\bk^\prime+2\sqrt{\gamma}
\widetilde{\bV}\right)\ \ .
\label{a21}
\end{equation}

\section{Born kernel}

In this appendix we show the invariant Born
kernel \eq{fm27} obtained by using the transformation
formula in Appendix A. These are functions
of $\bk^2$, $\bq^2$ and $\bk \cdot \bq$ with $\bk=\bq_f-\bq_i$,
$\bq=(\bq_f+\bq_i$)/2, and expressed by the special functions
given in \eq{a17}. Only $\theta$ with $\cos \theta=\widehat{\bq}_f
\cdot \widehat{\bq}_i$ is explicitly written in the
spatial functions $f^\Omega_\CT(\theta)$.
The $S^\prime$-type spatial
function $f^\Omega_{S^\prime}(\theta)$ is
obtained from $f^\Omega_S(\theta)$ with $\bk \rightarrow -\bk$.
There is no $E$-type possible for the non-central forces.
The partial wave decomposition of the Born
kernel is carried out numerically through \eq{fm28}.
The spin-flavor-color factors in the quark sector
are also shown in the operator form in the isospin space.

\subsection{Quark sector}

\noindent
\underline{Exchange normalization kernel}
\begin{equation}
M_N(\bq_f, \bq_i)=X_N\,f(\theta)\ \ ,
\label{b1}
\end{equation}
with
\begin{equation}
f(\theta)
=(\sqrt{3\pi} b)^3 \exp \left\{-{b^2 \over 3}(\bq^2+\bk^2)\right\}\ \ .
\label{b2}
\end{equation}

\bigskip

\noindent
\underline{Exchange kinetic-energy kernel}
\begin{equation}
M^K(\bq_f, \bq_i)
=X_N \left[{1 \over 3}\left(2+\li\right)+{1 \over 6}\left(1-\li\right)
Y\right]\,f_{K_1}(\theta)+X_K {1 \over 6}\left(1-\li\right)
\,f_{K_2}(\theta),
\label{b3}
\end{equation}
where $\lambda=(m_s/m_{ud})$, $Y$ is the hypercharge
of the total system, and
\begin{equation}
\left. \begin{array}{c}
f_{K_1}(\theta) \\
f_{K_2}(\theta) \\
\end{array} \right\}
={3 \over 4}x^2 m_{ud}c^2\,f(\theta)
\left\{ \begin{array}{c}
\left[-1+{1 \over 3}b^2(2\bq^2+\bk^2)\right] \\
{1 \over 4}\,\left[1+{2 \over 3}b^2(\bq^2-\bk^2)\right] \\
\end{array} \right.\ \ ,
\label{b4}
\end{equation}
with $x=(\hbar/m_{ud}cb)$.

\bigskip

\noindent
\underline{Color-Coulombic term}
\begin{equation}
M^{CC}(\bq_f, \bq_i)
=X_N \left[\,2\left(f^{CC}_E(\theta)-f^{CC}_S(\theta)
-f^{CC}_{S^\prime}(\theta)\right)+f^{CC}_{D_+}(\theta)
+f^{CC}_{D_-}(\theta)\right]\ \ ,
\label{b5}
\end{equation}
where
\begin{eqnarray}
& & f^{CC}_\CT(\theta)=\sqrt{{2 \over \pi}}\alpha_S x m_{ud}c^2
{4 \over 3}(\sqrt{3\pi} b)^3 \nonumber \\
& &  \times \left\{ \begin{array}{c}
\left({8 \over 11}\right)^{1 \over 2} \exp \left\{-{2 \over 11}
b^2 \left[{4 \over 3}(\bq^2+\bk^2)-\bk \cdot \bq \right]\right\}
~\widetilde{h}_0\left({1 \over \sqrt{11}}b |\bq+\bk|\right) \\
\left({1 \over 2}\right)^{1 \over 2} \exp \left\{-{1 \over 3}
b^2 \left(\bq^2+{1 \over 4} \bk^2 \right) \right\}
~\widetilde{h}_0\left({1 \over 2}b |\bk| \right) \\
\left({2 \over 3}\right)^{1 \over 2} \exp \left\{-{1 \over 3}
b^2 \bk^2 \right\}
~\widetilde{h}_0\left({1 \over \sqrt{3}} b |\bq| \right) \\
\end{array} \right.
~\hbox{for}~\CT= \left\{
\begin{array}{l}
S \\
D_+ \\
D_- \\
\end{array} \right. . \nonumber \\
\label{b6}
\end{eqnarray}
The $E$-type factor is given by $f^{CC}_E(\theta)
=\sqrt{2/\pi}\alpha_S x m_{ud}c^2\,(4/3)\,f(\theta)$.

\bigskip

\noindent
\underline{Breit retardation term}
\begin{eqnarray}
M^{MC}(\bq_f, \bq_i)
& = & X^{MC}_E \left(f^{MC}_E(\theta)
-{4 \over 9} f^{GC}_E(\theta)\right)
+X^{MC}_{S^\prime} \left(f^{MC}_{S^\prime}(\theta)
-{4 \over 9} f^{GC}_E(\theta)\right) \nonumber \\
& & +X^{MC}_S\,f^{MC}_S(\theta)
+X^{MC}_{D_+}\,f^{MC}_{D_+}(\theta)
+X^{MC}_{D_-}\,f^{MC}_{D_-}(\theta)\ \ ,
\label{b7}
\end{eqnarray}
where
\begin{eqnarray}
& & f^{MC}_\CT(\theta)=\sqrt{{2 \over \pi}}\alpha_S x^3 m_{ud}c^2
{2 \over 9}(\sqrt{3\pi} b)^3 \nonumber \\
& & \times \left\{ \begin{array}{l}
\exp \left\{-{1 \over 3}
b^2 (\bq^2+\bk^2)\right\} \left[{5 \over 2}-(b\bq)^2 \right]
\quad \hbox{for} \quad \CT=E \\
\left({8 \over 11}\right)^{1 \over 2} \exp \left\{-{2 \over 11}
b^2 \left[{4 \over 3}(\bq^2+\bk^2)-\bk \cdot \bq \right]\right\} \\
\qquad \times \left[ \widetilde{H}_0
\left({1 \over \sqrt{11}}b |\bq+\bk|\right)
-{3 \over 4}b^2(\bk \cdot \bq)
\widetilde{H}_1\left({1 \over \sqrt{11}}b |\bq+\bk|\right) \right. \\
\qquad \left. +{3 \over 22}b^4(\bq^2+\bk \cdot \bq) (\bk^2+\bk \cdot \bq)
\widetilde{H}_2\left({1 \over \sqrt{11}}b |\bq+\bk|\right) \right]
\quad \hbox{for} \quad \CT=S \\
\left({1 \over 2}\right)^{1 \over 2} \exp \left\{-{1 \over 3}
b^2 \left(\bq^2+{1 \over 4} \bk^2 \right) \right\}
\left[ -{1 \over 2} \widetilde{H}_0
\left({1 \over 2} b |\bk|\right)\right. \\ 
\qquad \left. +{3 \over 4}(b \bq)^2
\widetilde{H}_1\left({1 \over 2} b |\bk|\right)
-{3 \over 8} b^4 (\bk \cdot \bq)^2
\widetilde{H}_2\left({1 \over 2}b |\bk|\right) \right]
\quad \hbox{for} \quad \CT=D_+ \\
\left({2 \over 3}\right)^{1 \over 2} \exp \left\{-{1 \over 3}
b^2 \bk^2 \right\}
\left[ -2 \widetilde{H}_0\left({1 \over \sqrt{3}} b |\bq|\right)
+{3 \over 4}(b \bk)^2
\widetilde{H}_1\left({1 \over \sqrt{3}} b |\bq|\right) \right. \\
\qquad \left. -{1 \over 2} b^4 (\bk \cdot \bq)^2
\widetilde{H}_2\left({1 \over \sqrt{3}} b |\bq|\right) \right]
\quad \hbox{for} \quad \CT=D_-\ \ . \\
\end{array} \right.
\label{b8}
\end{eqnarray}
The functions $\widetilde{H}_n(x)$ with $n=0,~1,~2$ are
expressed by $\widetilde{h}_n(x)$ through
\begin{eqnarray}
\widetilde{H}_0(x)& = & \widetilde{h}_0(x) \nonumber \\
\widetilde{H}_1(x)& = & \widetilde{h}_0(x)
+{1 \over 3}\,\widetilde{h}_1(x) \nonumber \\
\widetilde{H}_2(x)& = & {1 \over 3}\,\widetilde{h}_1(x)
-{1 \over 5}\,\widetilde{h}_2(x)\ \ .
\label{b9}
\end{eqnarray}
The function $f^{GC}_E(\theta)$ is given below.

\bigskip

\noindent
\underline{Color-Magnetic term}
\begin{equation}
M^{GC}(\bq_f, \bq_i)
=-X^{GC}_{S^\prime}\,f^{GC}_E(\theta)+\sum_{\CT \neq E}
X^{GC}_{\CT}\,f^{GC}_{\CT}(\theta)\ \ ,
\label{b10}
\end{equation}
where $f^{GC}_E(\theta)
=\sqrt{2/\pi}\alpha_S x^3 m_{ud}c^2\,f(\theta)$ and
\begin{eqnarray}
& & f^{GC}_\CT(\theta)=\sqrt{{2 \over \pi}}\alpha_S x^3 m_{ud}c^2
(\sqrt{3\pi} b)^3 \nonumber \\
& &  \times \left\{ \begin{array}{c}
\left({8 \over 11}\right)^{3 \over 2} \exp \left\{-{2 \over 11}
b^2 \left[{4 \over 3}(\bq^2+\bk^2)-\bk \cdot \bq \right]\right\} \\
\left({1 \over 2}\right)^{3 \over 2} \exp \left\{-{1 \over 3}
b^2 \left(\bq^2+{1 \over 4} \bk^2 \right) \right\} \\
\left({2 \over 3}\right)^{3 \over 2} \exp \left\{-{1 \over 3}
b^2 \bk^2 \right\} \\
\end{array} \right.
~\hbox{for}~\CT= \left\{
\begin{array}{l}
S \\
D_+ \\
D_- \\
\end{array} \right.\ \ . \nonumber \\
\label{b11}
\end{eqnarray}

\bigskip

\noindent
\underline{$LS$ term}

In general there are three different types of $LS$ terms
in the $YN$ interaction;
i.e., $\Omega=LS$, $LS^{(-)}$ and $LS^{(-)}\sigma$. \cite{FU97}
These are different only for the spin-flavor-color factors.
For each type we have
\begin{equation}
M^{\Omega}(\bq_f, \bq_i)
=\sum_{\CT \neq E} X^{\Omega}_{\CT}\,f^{LS}_{\CT}(\theta)\ \ ,
\label{b12}
\end{equation}
where
\begin{eqnarray}
& & f^{LS}_\CT(\theta)=-\sqrt{{2 \over \pi}}\alpha_S x^3 m_{ud}c^2
{1 \over 2}(3\pi)^{3 \over 2} b^5 \nonumber \\
& &  \times \left\{ \begin{array}{c}
\left({8 \over 11}\right)^{3 \over 2} \exp \left\{-{2 \over 11}
b^2 \left[{4 \over 3}(\bq^2+\bk^2)-\bk \cdot \bq \right]\right\}
~\widetilde{h}_1\left({1 \over \sqrt{11}}b |\bq+\bk|\right) \\
\left({1 \over 2}\right)^{3 \over 2} \exp \left\{-{1 \over 3}
b^2 \left(\bq^2+{1 \over 4} \bk^2 \right) \right\}
~\widetilde{h}_1\left({1 \over 2}b |\bk| \right) \\
\left({2 \over 3}\right)^{3 \over 2} \exp \left\{-{1 \over 3}
b^2 \bk^2 \right\}
~\widetilde{h}_1\left({1 \over \sqrt{3}} b |\bq| \right) \\
\end{array} \right.
~\hbox{for}~\CT= \left\{
\begin{array}{l}
S \\
D_+ \\
D_- \\
\end{array} \right. . \nonumber \\
\label{b13}
\end{eqnarray}

\bigskip

\noindent
\underline{Tensor term}

There are three different types of tensor terms
as in \eq{fm26}. \cite{FU97}
These are given by
\begin{eqnarray}
\left. \begin{array}{c}
M^{T}(\bq_f, \bq_i) \\
M^{T^\prime}(\bq_f, \bq_i) \\
\end{array} \right\}
& = & X^T_{S}\,f^T_S(\theta)+X^T_{S^\prime}\,f^T_{S^\prime}(\theta)
+\left\{ \begin{array}{c}
X^T_{D_+}\,f^T_{D_+}(\theta) \\
X^T_{D_-}\,f^T_{D_-}(\theta) \\
\end{array} \right.\ \ , \nonumber \\
M^{T^{\prime \prime}}(\bq_f, \bq_i)
& = & 2 \left[\,X^T_{S}\,f^T_S(\theta)
-X^T_{S^\prime}\,f^T_{S^\prime}(\theta)\,\right]\ \ ,
\label{b14}
\end{eqnarray}
where
\begin{eqnarray}
& & f^T_\CT(\theta)=-\sqrt{{2 \over \pi}}\alpha_S x^3 m_{ud}c^2
{2 \over 5}(3\pi)^{3 \over 2} b^5 \nonumber \\
& &  \times \left\{ \begin{array}{c}
\left({8 \over 11}\right)^{3 \over 2}{1 \over 11}
\exp \left\{-{2 \over 11}
b^2 \left[{4 \over 3}(\bq^2+\bk^2)-\bk \cdot \bq \right]\right\}
~\widetilde{h}_2\left({1 \over \sqrt{11}}b |\bq+\bk|\right) \\
\left({1 \over 2}\right)^{3 \over 2}{1 \over 4}
\exp \left\{-{1 \over 3}
b^2 \left(\bq^2+{1 \over 4} \bk^2 \right) \right\}
~\widetilde{h}_2\left({1 \over 2}b |\bk| \right) \\
\left({2 \over 3}\right)^{3 \over 2}{1 \over 3}
\exp \left\{-{1 \over 3}
b^2 \bk^2 \right\}
~\widetilde{h}_2\left({1 \over \sqrt{3}} b |\bq| \right) \\
\end{array} \right.
~\hbox{for}~\CT= \left\{
\begin{array}{l}
S \\
D_+ \\
D_- \\
\end{array} \right. . \nonumber \\
\label{b15}
\end{eqnarray}

\subsection{EMEP sector}
 
%\bigskip

\noindent
\underline{Direct terms}

The S-meson central ($CN$), PS-meson spin-spin ($SN$) and PS-meson
tensor ($TN$) terms are given by
\begin{eqnarray}
M^{CN}_D(\bq_f, \bq_i)
& = & X^{CN}_{0D_+}\,f^{CN}_D(\theta)\ \ , \nonumber \\
M^{SN}_D(\bq_f, \bq_i)
& = & X^{T}_{0D_+} (\bfsigma_1 \cdot \bfsigma_2)
\,f^{SN}_D(\theta)\ \ , \nonumber \\
M^{TN}_D(\bq_f, \bq_i)
& = & X^{T}_{0D_+}
\,f^{TN}_D(\theta)\ \ ,
\label{b16}
\end{eqnarray}
where
\begin{eqnarray}
& & f^{CN}_D(\theta)=-4\pi \hbar c {1 \over m^2+\bk^2}
~e^{-{1 \over 3}(b \bk)^2}\ \ , \nonumber \\
& & f^{SN}_D(\theta)=4\pi \hbar c {1 \over 3}
\left( {1 \over m^2+\bk^2}-c_\delta {1 \over m^2}\right)
~e^{-{1 \over 3}(b \bk)^2}\ \ , \nonumber \\
& & f^{TN}_D(\theta)=-4\pi \hbar c {1 \over 3 m^2}
{1 \over m^2+\bk^2}~e^{-{1 \over 3}(b \bk)^2}
={1 \over 3 m^2}\,f^{CN}_D(\theta)\ \ .
\label{b17}
\end{eqnarray}
Here $m$ ($=mc/\hbar$) is the meson mass,
and the relationship $X^{SN}_{0D_+}=X^{T}_{0D_+}
(\bfsigma_1 \cdot \bfsigma_2)$ is used in \eq{b16}.
The factor $c_\delta$ in \eq{b17} is a reduction factor for
the spin-spin contact term of the PS-mesons. 

\bigskip

\noindent
\underline{Exchange terms}
\begin{eqnarray}
M^{({CN \atop SN})}(\bq_f, \bq_i) & = & -X^{({CN \atop SN})}_{1S^\prime}
\,f^{({CN \atop SN})}_E(\theta)
+\sum_{\CT \neq E} X^{({CN \atop SN})}_{1 \CT}
\,f^{({CN \atop SN})}_{\CT}(\theta)\ \ , \nonumber \\
\left. \begin{array}{c}
M^{TN}(\bq_f, \bq_i) \\
M^{TN^\prime}(\bq_f, \bq_i) \\
\end{array} \right\}
& = & X^T_{1S}\,f^{TN}_S(\theta)+X^T_{1S^\prime}
\,f^{TN}_{S^\prime}(\theta)
+\left\{ \begin{array}{c}
X^T_{1D_+}\,f^{TN}_{D_+}(\theta) \\
X^T_{1D_-}\,f^{TN}_{D_-}(\theta) \\
\end{array} \right.\ \ , \nonumber \\
M^{TN^{\prime \prime}}(\bq_f, \bq_i)
& = & 2 \left[\,X^T_{1S}\,f^{TN}_S(\theta)
-X^T_{1S^\prime}\,f^{TN}_{S^\prime}(\theta)\,\right]\ \ ,
\label{b18}
\end{eqnarray}
where
\begin{eqnarray}
& & \left. \begin{array}{c}
f^{CN}_\CT(\theta) \\
f^{SN}_\CT(\theta) \\
\end{array} \right\}=-4\pi
\left\{ \begin{array}{c}
1 \\ -{1 \over 3} \\
\end{array} \right\}
\left({3 \over 2} \right)^{{3 \over 2}} \hbar c b^2 \nonumber \\
& & \times \left\{ \begin{array}{c}
\exp \left\{-{1 \over 3}
b^2 (\bq^2+\bk^2)\right\} \left[\widetilde{\CY}_{\alpha_E}(0)
-\left\{ \begin{array}{c}
0 \\ c_\delta {1 \over 2\alpha_E} \\
\end{array} \right\} \right] \quad \hbox{for} \quad E \nonumber \\
\left({8 \over 11}\right)^{1 \over 2} \exp \left\{-{2 \over 11}
b^2 \left[{4 \over 3}(\bq^2+\bk^2)-\bk \cdot \bq \right]\right\}
\left[ \widetilde{\CY}_{\alpha_S}\left(
{1 \over \sqrt{11}} b |\bq+\bk| \right)
-\left\{ \begin{array}{c}
0 \\ c_\delta {1 \over 2\alpha_S} \\
\end{array} \right\} \right] \\
\hfill  \quad \hbox{for} \quad \CT=S \\
\left({1 \over 2}\right)^{1 \over 2} \exp \left\{-{1 \over 3}
b^2 \left(\bq^2+{1 \over 4} \bk^2 \right) \right\}
\left[ \widetilde{\CY}_{\alpha_{D_+}}\left({1 \over 2} b |\bk|\right)
-\left\{ \begin{array}{c}
0 \\ c_\delta {1 \over 2\alpha_{D_+}} \\
\end{array} \right\} \right] \quad \hbox{for} \quad \CT=D_+ \\
\left({2 \over 3}\right)^{1 \over 2} \exp \left\{-{1 \over 3}
b^2 \bk^2 \right\}
\left[ \widetilde{\CY}_{\alpha_{D_-}}\left({1 \over \sqrt{3}} b
|\bq|\right)-\left\{ \begin{array}{c}
0 \\ c_\delta {1 \over 2\alpha_{D_-}} \\
\end{array} \right\} \right] \quad \hbox{for} \quad \CT=D_-\ \ , \\
\end{array} \right. \nonumber \\
& & f^{TN}_\CT(\theta)=-4\pi \left({3 \over 2} \right)^{{1 \over 2}}
\hbar c b^2 \left({\hbar \over mc}\right)^2 \nonumber \\
& & \times \left\{ \begin{array}{c}
\left({8 \over 11}\right)^{3 \over 2}{1 \over 11}
\exp \left\{-{2 \over 11}
b^2 \left[{4 \over 3}(\bq^2+\bk^2)-\bk \cdot \bq \right]\right\}
~\widetilde{\CZ}_{\alpha_S}\left(
{1 \over \sqrt{11}} b |\bq+\bk| \right) \quad \hbox{for} \quad \CT=S \\
\left({1 \over 2}\right)^{3 \over 2}{1 \over 4}
\exp \left\{-{1 \over 3}
b^2 \left(\bq^2+{1 \over 4} \bk^2 \right) \right\}
~\widetilde{\CZ}_{\alpha_{D_+}}\left({1 \over 2} b |\bk|\right)
\quad \hbox{for} \quad \CT=D_+ \\
\left({2 \over 3}\right)^{3 \over 2}{1 \over 3}
\exp \left\{-{1 \over 3} b^2 \bk^2 \right\}
~\widetilde{\CZ}_{\alpha_{D_-}}\left({1 \over \sqrt{3}} b
|\bq|\right) \quad \hbox{for} \quad \CT=D_-\ \ . \\
\end{array} \right.
\label{b19}
\end{eqnarray}
The coefficients $\alpha_\CT$ are given
by $\alpha_S=\alpha_{S^\prime}=(11/8) \alpha_E$,
$\alpha_{D_+}=2\alpha_E$, and $\alpha_{D_-}=(3/2)\alpha_E$,
with $\alpha_E=(mb)^2/2=(1/2)(mcb/\hbar)^2$.

\subsection{Spin-flavor-color factors in the quark sector}

The spin-flavor-color factors in the quark sector
for the $NN$ system are given in Eq.\,(B.6) of Ref.\,\citen{FU87}.
We should note that $X_\CT$ there is
actually $X^{GC}_\CT$ for $\CT=D_+$ and $D_-$,
and $X^{GC}_S=X^{GC}_{S^\prime}=-X_E$.
(Note that $X_E$ is not $X^{GC}_E$.)
Furthermore, $X^{MC}_{D_-}=X^{MC}_{D_+}=X_N$, $X^{MC}_S
=X^{MC}_{S^\prime}=-2 X_N$, $X^{MC}_E=-X^N$, and $X_K=0$
for $NN$.
The spin-orbit and tensor factors for $YN$ systems
are found in Appendix C of Ref.\,\citen{FU97}.
Here we list up the one-quark exchange ($x=1$) central
factors $(X^\Omega_\CT)_{B_3 B_1}$ in \eq{a3},
with respect to the quark sector of the $\Lambda N$,
$\Sigma N$ and $\Xi N$ systems.
The spin-flavor factors in the EMEP sector
and the detailed derivation of these factors
will be published elsewhere.
In the following,
the parameter $\lambda$ ($=m_s/m_{ud}$\footnote{$\lambda=m_{ud}/m_s$ in
Ref.\,\protect\citen{FU97} is a misprint.}) controls
the flavor symmetry breaking of the FB interaction.
The spin and isospin operators, $\bfsigma_i$ and $\bftau_i$, are
with respect to the two baryons $B_i$ with $i=1$ or 2 in the
initial state, and the flavor exchange operator $P_F$ is supposed
to operate on the ket state.
We assume $B_2=B_4=N$.
The spin-flavor-color factor for the exchange normalization kernel
is given by $X_N=X^{MC}_{D_-}(\lambda=1)$.
The factors of the interaction type $\CT=S^\prime$ are discussed
in $\S$ 5 of Ref.\,\citen{FU97}. 

\bigskip

\noindent
[ $B_3 B_1=\Lambda \Lambda$ ]
\begin{eqnarray}
& & X_K=1-(1+\ssigma) P_F\ \ ,\nonumber \\
& & X^{MC}_{D_-}=-{1 \over 4} \left[ 2+\lam{1}\left(1+\ssigma \right)
P_F \right] \ \ ,\nonumber \\
& & X^{MC}_{D_+}=-{1 \over 4} \left[ 1+\lam{1}
+\left(1+\ssigma \right) P_F \right] \ \ ,\nonumber \\
& & X^{MC}_S={1 \over 4} \left[ 3+\lam{1}
+\left(1+\lam{1}\right)\left(1+\ssigma \right) P_F \right]
\ \ ,\nonumber \\
& & \qquad \nonumber \\
& & X^{GC}_{D_-}={1 \over 2}+{1 \over 24}
\left[ \left(1+\lam{4}+\lab{1}\right)
+\left(\ssigma\right) \left(1-{4 \over 3 \lambda}
+\lab{1}\right) \right] P_F\ \ ,\nonumber \\
& & X^{GC}_{D_+}={1 \over 72} \left[ \left(13+\lab{3}\right)
+6 P_F+\left(\ssigma\right)\left({4 \over 3 \lambda}+6 P_F\right)
\right]\ \ ,\nonumber \\
& & X^{GC}_S=-{1 \over 72} \left\{ \left(1+\lab{3}\right)
+3 \left(3+\lab{1}\right) P_F+\left(\ssigma\right)
\left[ {4 \over 3 \lambda}+3 \left(3+\lab{1}\right) P_F
\right] \right\}\ \ . \nonumber \\
& & \qquad
\label{b20}
\end{eqnarray}
%
%
%\bigskip

\noindent
[ $B_3 B_1=\Sigma \Sigma$ ]
\begin{eqnarray}
& & X_K=1+{1 \over 3} \ttau -{1 \over 3} \ttaup
+\left(\ssigma\right){2 \over 9}\left[1+{5 \over 3}\ttau
-{1 \over 6}\ttaup \right]\ \ ,\nonumber \\
& & X^{MC}_{D_-}=-{1 \over 2} \left\{1+{1 \over 3}\ttau
+{1 \over 6\lambda}\ttaup \right. \nonumber \\
& & \qquad \quad \left. +\left(\ssigma\right){2 \over 9}
\left[1+{5 \over 3}\ttau +{1 \over 12 \lambda}\ttaup \right]
\right\}\ \ ,\nonumber \\
& & X^{MC}_{D_+}=-{1 \over 4}\left(1+\lam{1}\right)
\left[ 1+{1 \over 3}\ttau +\left(\ssigma\right){2 \over 9}
\left(1+{5 \over 3}\ttau \right) \right] \nonumber \\
& & \qquad \quad -{1 \over 12}
\left(1+{1 \over9}\ssigma\right)\ttaup\ \ ,\nonumber \\
& & \qquad  \nonumber \\
& & X^{MC}_S={1 \over 4}\left(3+\lam{1}\right)
\left[ 1+{1 \over 3}\ttau +\left(\ssigma\right){2 \over 9}
\left(1+{5 \over 3}\ttau \right) \right] \nonumber \\
& & \qquad \quad +{1 \over 12}\left(1+\lam{1}\right)
\left(1+{1 \over9}\ssigma\right)\ttaup\ \ ,\nonumber \\
& & \qquad  \nonumber \\
& & X^{GC}_{D_-}={1 \over 2} \left(1+{1 \over 3}\ttau \right)
+{1 \over 72}\left(1+\lam{4}+\lab{1}\right)\ttaup \nonumber \\
& & \qquad \quad +\left(\ssigma\right){1 \over 81}
\left[1+{5 \over 3}\ttau +{1 \over 8}\left(1-{4 \over 3\lambda}
+\lab{1}\right)\ttaup \right]\ \ ,\nonumber \\
& & X^{GC}_{D_+}={1 \over 3 \cdot 72} \left\{
\left(23+\lam{8}+\lab{9}\right)+{1 \over 3}\left(31-\lam{8}
+\lab{9}\right)\ttau 
+{34 \over 3}\ttaup \right. \nonumber \\
& & \quad \quad \left. +\left(\ssigma\right){2 \over 3}\left[
\left(13-\lam{2}+\lab{3}\right)+{1 \over 3}\left(41+\lam{2}
+\lab{15}\right)\ttau+{11 \over 3}\ttaup \right] \right\}, \nonumber \\ 
& & X^{GC}_S=-{1 \over 72} \left\{
\left(17-\lam{8}+\lab{3}\right)+\left(11-{8 \over 3 \lambda}
+\lab{1}\right)\ttau \right.\nonumber \\
& & \qquad \quad 
+{1 \over 3}\left(1-\lam{8}+\lab{3}\right) \ttaup
+\left(\ssigma\right){1 \over 9}\left[
6 \left(11-{2 \over 3 \lambda}+\lab{1}\right)
\right. \nonumber \\
& & \qquad \quad \left. \left.
+2\left(39-{10 \over 3 \lambda}+\lab{5}\right)\ttau
+{1 \over 3}\left(1-\lam{8}+\lab{3}\right) \ttaup \right] \right\}
\ \ .
\label{b21}
\end{eqnarray}
%
%
%\bigskip

\noindent
[ $B_3 B_1=\Xi \Xi$ ]
\begin{eqnarray}
& & X_K=1+{1 \over 3} \ttau -\left(\ssigma\right){1 \over 9}
\left(1+{5 \over 3}\ttau\right)\ \ ,\qquad
X_N=-{1 \over 4} X_K\ \ ,\nonumber \\
& & X^{MC}_{D_-}=X_N\ \ ,\qquad
X^{MC}_{D_+}=\lam{1} X_N\ \ ,\qquad 
X^{MC}_S=-\left(1+\lam{1}\right) X_N\ \ ,\nonumber \\
& & \qquad  \nonumber \\
& & X^{GC}_{D_-}={1 \over 4}\left[
1+{1 \over 3} \ttau -\left(\ssigma\right){1 \over 81}
\left(1+{5 \over 3}\ttau\right) \right]\ \ ,\nonumber \\
& & X^{GC}_{D_+}={1 \over 3 \cdot 72} \left\{
\left(9+\lam{8}+\lab{9}\right)
+{1 \over 3}\left(9-\lam{8}+\lab{9}\right)\ttau \right. \nonumber \\
& & \qquad \quad \left. -\left(\ssigma\right){1 \over 3}
\left[ \left(3-\lam{8}+\lab{3}\right)
+{1 \over 3}\left(15+\lam{8}+\lab{15}\right) \ttau \right] \right\}
\ \ ,\nonumber \\
& & X^{GC}_S=-{1 \over 72} \left\{
\left(5-\lam{8}+\lab{3}\right)
+{1 \over 3}\left(13-\lam{8}+\lab{3}\right)\ttau \right. \nonumber \\
& & \qquad \left. -\left(\ssigma\right){1 \over 9}
\left[ \left(13-\lam{8}+\lab{3}\right)
+{1 \over 3}\left(41-\lam{40}+\lab{15}\right) \ttau \right] \right\}\ .
\label{b22}
\end{eqnarray}
%

%\bigskip
\noindent
[ $B_3 B_1=\Lambda \Sigma$ and $\Sigma \Lambda$ ]
\begin{eqnarray}
& & X_K=P_F-\left(\ssigma\right){1 \over 3}
\left({5 \over 3}+P_F\right)\ \ ,\nonumber \\
& & X^{MC}_{D_-}={1 \over 4} \left[\lam{1} P_F
+\left(\ssigma\right){1 \over 3}
\left({10 \over 3}-\lam{1} P_F \right) \right]\ \ ,\nonumber \\
& & X^{MC}_{D_+}={5 \over 36}\left(1+\lam{1}\right)
\left(\ssigma\right)+{1 \over 4}\left(1-{1 \over 3}\ssigma \right)
P_F\ \ ,\nonumber \\
& & X^{MC}_S=-{5 \over 36}\left(3+\lam{1}\right)
\left(\ssigma\right)-{1 \over 4}\left(1+\lam{1}\right)
\left(1-{1 \over 3}\ssigma \right) P_F\ \ ,\nonumber \\
& & \qquad  \nonumber \\
& & X^{GC}_{D_-}=-{5 \over 2 \cdot 81}\left(\ssigma\right)
-{1 \over 24}\left[\left(1+\lam{4}+\lab{1}\right)
-\left(\ssigma\right){1 \over 3}\left(1-{4 \over 3 \lambda}
+\lab{1}\right) \right] P_F\ \ ,\nonumber \\
& & X^{GC}_{D_+}=-{1 \over 3 \cdot 36} \left[
\left(\lam{2}+9 P_F\right)+\left(\ssigma\right){1 \over 6}
\left(49+\lab{15}-18 P_F\right) \right]\ \ ,\nonumber \\
& & X^{GC}_S=\left\{
\begin{array}{l}
\dfrac{1}{72} \left\{ \dfrac{4}{\lambda}
+\left(9-\dfrac{8}{\lambda}+\dfrac{3}{\lambda^2} \right) P_F
\right. \\
\qquad \left. +\left(\ssigma\right)\dfrac{1}{3}
\left[ \left(39-\dfrac{40}{3\lambda}+\dfrac{5}{\lambda^2} \right)
-\left(9-\dfrac{8}{\lambda}+\dfrac{3}{\lambda^2} \right) P_F
\right] \right\} \\
\dfrac{1}{72} \left\{ \dfrac{4}{\lambda}
+\left(1+\dfrac{3}{\lambda^2} \right) P_F
\right. \\
\qquad \left. +\left(\ssigma\right)\dfrac{1}{3}
\left[ \left(\dfrac{37}{3}+\dfrac{40}{3\lambda}
+\dfrac{5}{\lambda^2} \right)
-\left(1+\dfrac{3}{\lambda^2} \right) P_F
\right] \right\} \\
\end{array} \right. \nonumber \\
& & \hspace{95mm} \hbox{for} \quad \left\{
\begin{array}{c}
\Lambda \Sigma \\ [5mm]
\Sigma \Lambda \\
\end{array} \right.\ .
\label{b23}
\end{eqnarray}

\newpage

\section{Partial-wave decomposition of the Born kernel}

In this appendix we derive some formulae for the partial-wave
decomposition of the Born kernel.
Suppose the basic Born kernel \eq{fm23} is expanded
into partial waves as
\begin{eqnarray}
& & M(\bq_f, \bq_i)=\sum_\Omega M^\Omega(\bq_f, \bq_i)
\,\CO^\Omega(\bq_f, \bq_i) \nonumber \\
& & = \sum_{JM\ell \ell^\prime S S^\prime}
4\pi~R^J_{S^\prime \ell^\prime, S \ell}(q_f, q_i)
~\CY_{(\ell^\prime S^\prime)JM}(\widehat{\bq}_f; spin)
\CY^*_{(\ell S)JM}(\widehat{\bq}_i; spin)\ \ ,
\label{c1}
\end{eqnarray}
where $\CY_{(\ell S)JM}(\widehat{\bq}; spin)=[Y_\ell(\widetilde{\bq})
\chi_S(spin)]_{JM}$ is the standard space-spin function. 
We can show that the partial-wave components are given by
\begin{eqnarray}
R^J_{S^\prime \ell^\prime, S \ell}(q_f, q_i) & = &
\sum_\Omega R^{\Omega J}_{S^\prime \ell^\prime, S \ell}
(q_f, q_i)\ \ , \nonumber \\
R^{\Omega J}_{S^\prime \ell^\prime, S \ell}
(q_f, q_i) & = & {1 \over 2}\int_{-1}^1
M^\Omega(\bq_f, \bq_i)\,I^{\Omega J}_{S^\prime \ell^\prime, S \ell}
(\theta)\,d\,(\cos \theta)\ \ ,
\label{c2}
\end{eqnarray}
with $\cos \theta=(\widehat{\bq}_f \cdot \widehat{\bq}_i)$ and  
\begin{equation}
I^{\Omega J}_{S^\prime \ell^\prime, S \ell}
(\theta)={4\pi \over 2J+1} \sum_M \langle
\CY_{(\ell^\prime S^\prime)JM}(\widehat{\bq}_f; spin)
|\CO^\Omega(\bq_f, \bq_i)|\CY_{(\ell S)JM}(\widehat{\bq}_i; spin)
\rangle_{spin} \ \ ,
\label{c3}
\end{equation}
by using the rotational invariance
of $R^{\Omega J}_{S^\prime \ell^\prime, S \ell}(q_f, q_i)$.
In \eq{c3} the matrix elements are with respect to the spin variables.
For the spin-independent central ($\CO^{CC}=1$) and
the spin-spin ($\CO^{SS}=(\bfsigma_1 \cdot \bfsigma_2)$) terms,
the factors \eq{c3} are given by
\begin{equation}
I^{\left\{{CC \atop SS}\right\} J}_{S^\prime \ell^\prime, S \ell}
(\theta)=\delta_{S^\prime, S} \delta_{\ell^\prime, \ell}
\left\{ \begin{array}{c}
1 \\
\left[2S(S+1)-3\right] \\
\end{array} \right\}
P_\ell(\cos \theta)\ \ .
\label{c4}
\end{equation}
For the $LS$ and tensor terms
with Eqs.\,(\ref{fm24}) and (\ref{fm26}),
some calculations using the angular-momentum algebra yield
\begin{eqnarray}
I^{LS~J}_{S^\prime \ell^\prime, S \ell}(\theta)
& = & -\delta_{S, 1} \delta_{S^\prime, S} \delta_{\ell^\prime, \ell}
q_f q_i {1 \over 2} \left[ 1+{2-J(J+1) \over \ell (\ell+1)} \right]
\nonumber \\
& & \times \sin \theta\,P^1_\ell(\cos \theta)
\quad \hbox{for} \quad \ell=J,~J \pm 1\ \ ,\nonumber \\
I^{\left\{ {LS^{(-)} \atop LS^{(-)}\sigma} \right\} J}
_{S^\prime \ell^\prime, S \ell}(\theta)
& = & \delta_{J, \ell} \delta_{\ell^\prime, \ell}
\left\{ \begin{array}{c}
1 \\
\left[S(S+1)-1\right] \\
\end{array} \right\}
q_f q_i {1 \over \sqrt{J(J+1)}} \sin \theta\,P^1_J(\cos \theta)
\nonumber \\
& & \quad \hbox{for} \quad S^\prime,
~S=0,~1~~\hbox{or}~~1,~0\ \ ,\nonumber \\
I^{\left\{ {T \atop T^\prime} \right\} J}
_{1\ell^\prime, 1\ell}(\theta)
& = & 
\left\{ \begin{array}{c}
1 \\ {1 \over 4} \\
\end{array} \right\}
(S_{12})^J_{\ell^\prime, \ell}
\left[ {q_f}^2 P_{\ell}(\cos \theta)
+{q_i}^2 P_{\ell^\prime}(\cos \theta)
\right. \nonumber \\
& & \left.  \mp 2 q_f q_i \left\{
\begin{array}{c}
P_J(\cos \theta) \\
\cos \theta P_\ell(\cos \theta)
+{3 \over 2}{1 \over \ell(\ell+1)}\sin \theta P^1_\ell(\cos \theta) \\
\end{array} \right\}~\right] \nonumber \\
& & \ \ \hspace{25mm} \hbox{for} \quad \left\{
\begin{array}{c}
\ell^\prime=\ell \pm 2 \quad \hbox{and} \quad J=\ell \pm 1 \\
\ell=\ell^\prime=J,~J \pm 1 \\
\end{array} \right.\ \ ,\nonumber \\
I^{T^{\prime \prime}~J}_{1\ell^\prime, 1\ell}(\theta)
& = & {1\over 2}(S_{12})^J_{\ell^\prime, \ell}
\left[ {q_f}^2 P_{\ell}(\cos \theta)
-{q_i}^2 P_{\ell^\prime}(\cos \theta) \right]\ \ .
\label{c5}
\end{eqnarray}
Here $P^1_J(\cos\theta)=\sin\theta\,P^\prime_J(\cos\theta)$
is the associated Legendre function of the first kind with degree 1,
and $(S_{12})^J_{\ell^\prime, \ell}$ is the standard
tensor matrix elements.
The final braces in $I^{\left\{ {T/T^\prime} \right\} J}
_{1\ell^\prime, 1\ell}(\theta)$ in \eq{c5} are applied to
the choice of $\ell$ and $\ell^\prime$.
Note that the $LS$ and tensor components are non-zero only for
the spin triplet states,
while the $LS^{(-)}$ and $LS^{(-)}\sigma$ components are
accompanied with the spin transition between $S=0$ and 1.

The partial-wave components of the full Born kernel,
$V^{J~\Omega}_{S^\prime \ell^\prime, S \ell}(q_f, q_i)$ in
Eqs.\,(\ref{fm29}) and (\ref{fm30}), are easily obtained
from the above formulae.

\section{Partial-wave decomposition of invariant amplitudes}

Reconstruction of the invariant amplitudes in terms
of the $S$-matrices is an opposite process to the partial-wave
decomposition of the Born amplitudes, given in the preceding
appendix. We can use the same formulae,
Eqs.\,(\ref{c4}) and (\ref{c5}), to derive the
following result for the partial-wave decomposition of the
invariant amplitudes.

Suppose the invariant amplitudes \eq{in1} in the isospin basis
has the following decomposition:
\begin{eqnarray}
& & M_{ca}(\bq_f, \bq_i)=
\sqrt{(1+\delta_{c_1, c_2})(1+\delta_{a_1, a_2})}\nonumber \\
& & \times \sum_{JM\ell \ell^\prime S S^\prime}
4\pi~R^J_{\gamma S^\prime \ell^\prime, \alpha S \ell}
~\CY_{(\ell^\prime S^\prime)JM}(\widehat{\bq}_f; spin)
\CY^*_{(\ell S)JM}(\widehat{\bq}_i; spin)\ \ .
\label{d1}
\end{eqnarray}
Here $a=(a_1, a_2)$ and $c=(c_1, c_2)$ imply two-baryon configurations
corresponding to the initial and final states,
$\alpha$ and $\gamma$, respectively.
The front factor, $\sqrt{(1+\delta_{c_1, c_2})}$
$\times \sqrt{(1+\delta_{a_1, a_2})}$,
is a factor for two identical particles, and is 2 for $NN$,
for example. Without this trivial factor,
the invariant amplitudes for a fixed $c a$ are given by
\begin{eqnarray}
\left. \begin{array}{c}
g_0 \\
g_{ss} \\ 
\end{array}
\right \}
& = & {1\over 4}\,\sum_{J \ell S} (2J+1)
      \left\{ 
       \begin{array}{c}
              1 \\ 
           {1\over 3}[2S(S+1)-3] 
       \end{array}
      \right\} 
R^J_{S \ell, S \ell}~P_l(\cos\theta)\ \ , \nonumber \\
h_0 & = &  - {1\over 4} \sum_J 
       {(2J+1)\over J(J+1)}
\left[ R^J_{1 J, 1 J}~P^1_J(\cos\theta) \right.
\nonumber\\
& &  \left. + J~R^J_{1 J+1, 1 J+1} 
~P^1_{J+1}(\cos\theta)
- (J+1) R^J_{1 J-1, 1 J-1}
~P^1_{J-1}(\cos\theta) \right]\ \ , \nonumber \\
h_- & = & {1 \over 4} \sum_J 
{(2J+1) \over \sqrt{J(J+1)}}
\left[ R^J_{1 J, 0 J} + R^J_{0 J, 1 J}
\right]~P^1_J(\cos\theta)\ \ , \nonumber \\
f_- & = & - {1\over 4 }\sum_{J} {(2J+1)\over \sqrt{J(J+1)}}
\left[ R^J_{1 J, 0 J}
- R^J_{0 J, 1 J} \right]~P^1_J(\cos\theta)
\nonumber\ \ , \\
h_{k \choose P} & = & g_{ss} 
+h^{(0)} - {\cos \theta \over \sin \theta}
~h^{(1)} + {1 \over \sin \theta}
~h^{(3)} \nonumber \\
& & \mp \left\{ 
     \cos (\theta - 2 \phi)   
\left[ h^{(2)} - {\cos \theta \over \sin \theta}
~h^{(3)} + {1 \over \sin \theta}~h^{(1)}
\right] - \sin (\theta - 2 \phi)~f^{(1)} \right\}\ \ , \nonumber \\
h_n & = & g_{ss} - 2~\left[  
~h^{(0)} - {\cos \theta \over \sin \theta}
~h^{(1)} + {1 \over \sin \theta}~h^{(3)} \right]\ \ , \nonumber \\
f_+ & = & \sin (\theta - 2 \phi)~\left[
~h^{(2)} - {\cos \theta \over \sin \theta}
~h^{(3)} + {1 \over \sin \theta}~h^{(1)} \right]
+ \cos (\theta - 2 \phi)~f^{(1)} ,
\label{d2}  
\end{eqnarray}
with
\begin{eqnarray}
h^{(0)} & = & {1\over 6}\sum_{J} 
\left[ 
(2J+1)~R^J_{1 J, 1 J}~P_J(\cos\theta) 
-(J+2)~R^J_{1 J+1, 1 J+1}~P_{J+1}(\cos\theta)
\right. \nonumber\\
& & \left. -(J-1)~R^J_{1 J-1, 1 J-1}~P_{J-1}(\cos\theta) 
\right]\ \ ,\nonumber \\
h^{(1)} & = & {1\over 4}\sum_J 
\left[ {(2J+1)\over J(J+1)}~R^J_{1 J, 1 J} 
~P^1_J(\cos\theta) - {1\over (J+1)}~R^J_{1 J+1, 1 J+1} 
~P^1_{J+1}(\cos\theta)\right. \nonumber\\
& & \left. -{1\over J}~R^J_{1 J-1, 1 J-1}
~P^1_{J-1}(\cos\theta) \right]\ \ ,\nonumber \\
h^{(2)} & = & {1\over 2} \sum_J 
\sqrt{J(J+1)}~\left[ R^J_{1 J+1, 1 J-1}
+ R^J_{1 J-1, 1 J+1} \right]~P_J(\cos\theta)
\ \ ,\nonumber \\
h^{(3)} & = & {1\over 4} \sum_J {1\over \sqrt{J(J+1)}}
~\left[ R^J_{1 J+1, 1 J-1} 
+ R^J_{1 J-1, 1 J+1} \right] 
~P^1_J(\cos\theta)\ \ ,\nonumber \\
f^{(1)} & = & {1\over 4} \sum_J 
{(2J+1)\over \sqrt{J(J+1)}}~\left[ 
R^J_{1 J+1, 1 J-1} - R^J_{1 J-1, 1 J+1}
\right]~P^1_J(\cos\theta)\ \ .
\label{d3}  
\end{eqnarray}
In Eqs.\,(\ref{d2}) and (\ref{d3}) we have used shorthand notation
$R^J_{S^\prime \ell^\prime, S \ell}$ to denote
$R^J_{\gamma S^\prime \ell^\prime, \alpha S \ell}$,
where the spin values $S^\prime$ and $S$ are uniquely specified
from $\gamma$ and $\alpha$, respectively.
In \eq{d1} the summation over $\CP^\prime$ and $\CP$ is
already taken by adding up over all possible
values of $\ell$ and $S$,
for which $\CP$ is uniquely determined through the generalized
Pauli principle, $(-1)^\ell (-1)^{1-S} \CP=-1$.
Note that a simple relationship, $h_k+h_P+h_n= 3 g_{ss}$, holds.
The on-shell scattering amplitudes
are conveniently expressed by only two
angles $\theta$ and $\phi$, the latter of which is defined
through $\cos \phi=\widehat{\bP}\cdot \widehat{\bq}_i$ with
\begin{equation}
\cot\phi={\sin\theta \over \vert \bq_i \vert / \vert \bq_f \vert
- \cos\theta}\ \ .
\label{d4}
\end{equation}
For the elastic scattering with $a_1=c_1$ and $a_2=c_2$,
the magnitude of the c.m. momentum is equal before and after
the scattering ($\vert \bq_i \vert = \vert \bq_f \vert$),
thus yielding $\phi=\theta/2$.
In this case, the time-reversal invariance gives a further
simplification for non-zero invariant amplitudes.
Since the $S$-matrix is symmetric, $f_-$ in \eq{d2} and
$f^{(1)}$ in \eq{d3} are both zero. Combining these two
conditions, we find that $f_-=f_+=0$ for the elastic
scattering. This is the well-known result of six independent
invariant amplitudes for the elastic scattering
with the time-reversal invariance.


\newpage

\begin{thebibliography}{99}
%
\bibitem{CL79}
F. E. Close, {\em An Introduction to Quarks and Partons}
(Academic, London, 1979).
%
\bibitem{WS84} M. Oka and K. Yazaki,
in {\em Quarks and Nuclei}, ed. W. Weise
(World Scientific, Singapore, 1984), p.489;
K. Shimizu, Rep. Prog. Phys. {\bf52} (1989), 1;
C. W. Wong, Phys. Rep. {\bf 136} (1986), 1.
%
\bibitem{NA95} C. Nakamoto, Y. Suzuki and Y. Fujiwara,
Prog. Theor. Phys. {\bf 94} (1995), 65.
%
\bibitem{FU95} Y. Fujiwara, C. Nakamoto and Y. Suzuki,
Prog. Theor. Phys. {\bf 94} (1995) 215, 353.
%
\bibitem{FU96a} Y. Fujiwara, C. Nakamoto and Y. Suzuki,
Phys. Rev. Lett. {\bf 76} (1996), 2242.
%
\bibitem{FU96b} Y. Fujiwara, C. Nakamoto and Y. Suzuki,
Phys. Rev. {\bf C54} (1996), 2180.
%
\bibitem{FJ98} T. Fujita, Y. Fujiwara, C. Nakamoto
and Y. Suzuki, Prog. Theor. Phys. {\bf 100} (1998), 931.
%
\bibitem{YU95} Y. W. Yu, Z. Y. Zhang, P. N. Shen
and L. R. Dai, Phys. Rev. {\bf C52} (1995), 3393.
%
\bibitem{ZH97} Z. Y. Zhang, Y. W. Yu, P. N. Shen, L. R. Dai,
Amand Faessler and U. Straub, Nucl. Phys. {\bf A625} (1997), 59.
%
\bibitem{YA98} S. Yang, P. N. Shen, Z. Y. Zhang and Y. W. Yu,
Nucl. Phys. {\bf A635} (1998), 146.
%
\bibitem{KA77} M. Kamimura, Prog. Theor. Phys. Suppl. No. 62
(1977) 236.
%
\bibitem{KU93} K. Kume and S. Yamaguchi,
Phys. Rev. {\bf C48} (1993), 2097.
%
\bibitem{NO65} H. P. Noyes, Phys. Rev. Lett. {\bf 15} (1965), 538.
%
\bibitem{KO65} K.L. Kowalski, Phys. Rev. Lett. {\bf 15} (1965), 798.
[Erratum {\bf 15} (1965), 908.]
%
\bibitem{KO99} M. Kohno, Y. Fujiwara, T. Fujita,
C. Nakamoto and Y. Suzuki,
submitted to Nucl. Phys. A (1999).
%
\bibitem{SU84} Y. Suzuki, Nucl. Phys. {\bf A430} (1984), 539.
%
\bibitem{FU97} Y. Fujiwara, C. Nakamoto, Y. Suzuki
and Zhang Zong-ye,
Prog. Theor. Phys. {\bf 97} (1997), 587.
%
\bibitem{HT70} M. I. Haftel and F. Tabakin,
Nucl. Phys. {\bf A158} (1970),1.
%
\bibitem{BL52} J. M. Blatt, and L. C. Biedenharn,
Rev. Mod. Phys. {\bf 24} (1952), 258.
%
\bibitem{BJ76} G. E. Brown and A. D. Jackson, {\em The Nucleon-Nucleon
Interaction} (North Holland, Amsterdam, 1976).
%
\bibitem{SUP77} K. Ikeda, R. Tamagaki, S. Saito, H. Horiuchi,
A. Tohsaki-Suzuki and M. Kamimura,
Prog. Theor. Phys. Suppl. No. 62 (1977), 1. 
%
\bibitem{SU83} Y. Suzuki, Nucl. Phys. {\bf A405} (1983), 40.
%
\bibitem{KI94} Y. Fujiwara and Y. C. Tang,
Memoirs of the Faculty of Science, Kyoto University,
Series A of Physics, Astrophysics, Geophysics and Chemistry,
Vol. XXXIX, No. 1, Article 5 (1994), 91.
%
\bibitem{FU86} Y. Fujiwara and K. T. Hecht,
Nucl. Phys. {\bf A456} (1986), 669.
%
\bibitem{FU87} Y. Fujiwara and K. T. Hecht,
Nucl. Phys. {\bf A462} (1987), 621.
%
\bibitem{FU87s} Y. Fujiwara,
Prog. Theor. Phys. Suppl. No. 91 (1987), 160.
%
\bibitem{FU92} Y. Fujiwara,
Prog. Theor. Phys. {\bf 88} (1992), 933. 
%
\bibitem{SCHU} H.-J. Schulze, M. Baldo, U. Lombardo, J. Cugnon and
A. Lejeune, Phys. Rev. {\bf C57} (1998), 704.
%
\bibitem{SAID} R. A. Arndt, Scattering Analysis Interactive
Dial-up (SAID), Virginia Polytechnic Institute,
Blacksburg, Virginia (private communication).
%
\bibitem{PARI} M. Lacombe, B. Loiseau, J. M. Richard, R. Vinh Mau,
J. C\^{o}t\'{e}, P. Pir\`{e}s and R. de Tourreil,
Phys. Rev. {\bf C21} (1980), 861.
%
\bibitem{SP82} R. A. Arndt, L. D. Roper, R. A. Bryan, R. B. Clark,
B. J. VerWest and P. Signell, Phys. Rev. {\bf D28} (1983), 97. 
%
\bibitem{BR67} R. A. Bryan and B. L. Scott, Phys. Rev. {\bf 164}
(1967), 1215.
%
\bibitem{NSC89}
P. M. M. Maessen, Th. A. Rijken and J. J. de Swart, Phys. Rev.
{\bf C40} (1989), 2226.
%
\bibitem{NSC97}
Th. A. Rijken, V. G. J. Stoks and Y. Yamamoto, Phys. Rev.
{\bf C59} (1999), 21.
%
\bibitem{LS99} Y. Fujiwara, M. Kohno, T. Fujita,
C. Nakamoto and Y. Suzuki, submitted to Nucl. Phys. A (1999).
%
\end{thebibliography}
\end{document}